\documentclass[11pt,letterpaper]{article}
\usepackage[margin=1in]{geometry}
\usepackage[T1]{fontenc}
\usepackage[utf8]{inputenc}
\usepackage{enumerate,url}
\usepackage{amsmath,amsfonts,amssymb,amsthm,bm}
\usepackage{graphicx,tabularx,multicol,multirow,booktabs,array}
\usepackage{subcaption,longtable,placeins,xcolor,tcolorbox}
\usepackage[round,authoryear]{natbib}
\setcitestyle{aysep={}}
\usepackage{xurl}
\usepackage[colorlinks=true,allcolors=blue]{hyperref}
\definecolor{ExampleBlue}{RGB}{14,65,156}
\graphicspath{{figures/}}
\newcommand{\E}{\mathbb{E}}
\newcommand{\ind}{\mathbf{1}}

\newcommand{\logit}{\operatorname{logit}}

\theoremstyle{remark}
\newtheorem{remark}{Remark}

\newcommand{\RPStrongAccuracy}{97.30}

\newcommand{\RPLPAccuracy}{96.42}

\newcommand{\RPLPSaving}{20.38}

\newcommand{\RPEmbeddingKnnAccuracy}{96.41}

\newcommand{\RPEmbeddingKnnSaving}{43.95}

\newcommand{\RPEmbeddingRiskAccuracy}{96.42}

\newcommand{\RPEmbeddingRiskSaving}{53.39}

\newcommand{\RPLdaKnnAccuracy}{96.27}

\newcommand{\RPLdaKnnSaving}{26.98}

\newcommand{\RPLdaRiskAccuracy}{96.40}

\newcommand{\RPLdaRiskSaving}{45.19}

\newcommand{\RPEmbeddingReduction}{16.83}

\newcommand{\RPLdaReduction}{24.94}

\title{\textbf{Right, Wrong, or Unfinished: A Competing Risks Framework for Language-Model Reasoning under Token Budgets}}
\author{Boen Jiang, Guanhua Fang, and Wen Yu\\
\small Department of Statistics and Data Science, Fudan University}
\date{}
\hypersetup{pdftitle={Right, Wrong, or Unfinished: A Competing Risks Framework for Language-Model Reasoning under Token Budgets},pdfauthor={Boen Jiang, Guanhua Fang, Wen Yu}}

\begin{document}
\maketitle

% ===== BEGIN abstract.tex =====
\begin{abstract}
Assessing language-model performance under a token budget requires understanding both when a response will finish and whether it will be correct. 
We develop {\bf C}ompeting {\bf R}isks {\bf A}nalysis of {\bf T}oken {\bf E}xpenditure in {\bf R}easoning (CRATER), a statistically principled framework that treats correct and incorrect completion as distinct competing event types and responses unfinished at the generation limit as administratively right-censored.
The fitted distribution provides probabilities of correct completion, incorrect completion, and remaining unfinished across token budgets, together with expected token consumption. 
We evaluate the proposed method using responses to MATH questions and
demonstrate its value in two downstream applications.
For routing before generation, CRATER guides the joint choice
of a model configuration and token budget. Under hypothetical
token prices, it reduces mean cost by \RPEmbeddingRiskSaving\%
relative to a strong reference, with observed correct-completion
rates of \RPEmbeddingRiskAccuracy\% and \RPStrongAccuracy\%,
respectively.
For repeated sampling, we combine CRATER's single-attempt
success probabilities with an existing beta-binomial
calibration approach to predict pass@$k$.
Across both applications, CRATER delivers predictions at
any specified token budget up to the generation limit
from a common model of completion time and correctness,
eliminating the need for separate parameter estimation for each budget.
\end{abstract}
% ===== END abstract.tex =====


% ===== BEGIN MANUSCRIPT BODY =====
% ===== BEGIN introduction.tex =====
\section{Introduction}
\label{sec:introduction}

The usefulness of a language-model response depends on both its content and the resources required to produce it. Practical applications therefore require attention to the cost of obtaining a successful response as well as the probability of success. For example, FrugalGPT studies this trade-off in news classification, reading comprehension, and scientific question answering, showing that adaptive use of models can reduce expenditure while maintaining task accuracy \citep{chen2023frugal}. % <!--ref:chen2023frugal--><!--anchor:section:Abstract-->
In mathematical reasoning, \citet{snell2024scaling} show that the effective allocation of inference-time computation depends on problem difficulty. % <!--ref:snell2024scaling--><!--anchor:section:Abstract-->
These examples motivate decisions about which model to use, how long to let it generate, and how many attempts to allow. Such decisions benefit from predictions of both the probability of obtaining a correct answer within a specified budget and the resources that generation would consume.

Using predicted success probabilities to guide resource allocation requires understanding how well they reflect observed outcomes. Calibration, which asks whether predicted probabilities of correctness agree with observed frequencies, forms part of the broader statistical study of uncertainty in language-model outputs \citep{ji2025overview,tian2023calibration}. % <!--ref:ji2025overview--><!--anchor:section:4.1,6.1--><!--ref:tian2023calibration--><!--anchor:section:Abstract-->
\citet{kadavath2022know} study probabilities that a proposed answer is correct and that a model can answer a question correctly before a particular answer is given. % <!--ref:kadavath2022know--><!--anchor:section:Abstract-->
\citet{tian2023calibration} find that elicited verbal confidence can be better calibrated than model probabilities on question-answering benchmarks. % <!--ref:tian2023calibration--><!--anchor:section:Abstract-->
For resource allocation, these assessments of correctness must be considered together with the computation required to obtain a response. The relevant prediction is then the probability of correct completion within a specified budget, together with the expected resources consumed.

Output-length modeling addresses the resource component of this problem. For interactive serving, \citet{qiu2024length} predict output sequence lengths from input queries using a lightweight proxy model and use these predictions to prioritize shorter requests, reducing average request completion times and improving throughput. % <!--ref:qiu2024length--><!--anchor:section:3-->
During generation, \citet{piotrowski2025tokens} use hidden states from multiple transformer layers to predict the number of tokens remaining in a response. % <!--ref:piotrowski2025tokens--><!--anchor:section:2-->
Moving from length predictions to distributional modeling, \citet{zheng2026uncertainlength} fit log-$t$ distributions to output lengths and use information about their tails to improve scheduling for online inference and offline data generation. % <!--ref:zheng2026uncertainlength--><!--anchor:section:3,4,6-->
Model routing uses predictions of response quality and cost to guide model selection. CARROT estimates both quantities conditional on the input question, while R2-Router jointly selects a model and an output-length budget \citep{somerstep2025carrot,xue2026r2}. % <!--ref:somerstep2025carrot--><!--anchor:section:Abstract--><!--ref:xue2026r2--><!--anchor:section:Abstract-->
Together, these studies motivate probability assessments that account for both response quality and resource consumption. Our focus is a joint distribution that connects completion length to correctness, so that success probabilities and expected consumption at different budgets arise from a common statistical model.

At any token budget, a response can have completed correctly, completed incorrectly, or remain unfinished. Distinguishing these states matters because accuracy among completed responses conditions on a population that changes with the budget, while an unfinished response has an unknown eventual outcome. If observation stops at a generation limit before completion is recorded, its completion time is administratively right-censored. \citet{nagpal2025length} applies the Kaplan-Meier estimator to language-model generation lengths under right censoring. % <!--ref:nagpal2025length--><!--anchor:section:2-->
To incorporate correctness, we treat correct and incorrect completion as two competing event types, with unfinished responses remaining at risk. The competing-risks framework relates their cause-specific hazards to the probabilities of correct completion, incorrect completion, and no completion by each budget \citep{putter2007}. % <!--ref:putter2007--><!--anchor:section:Abstract-->
It thus retains information from capped responses while allowing the timing of the two completion types to differ. The target is the joint behavior of completion time and outcome within the observed token window, conditional on question information available before generation.

We propose \textbf{C}ompeting \textbf{R}isks \textbf{A}nalysis
of \textbf{T}oken \textbf{E}xpenditure in \textbf{R}easoning
(CRATER), a competing-risks based working model on the output-token
scale. CRATER combines question-embedding features with
configuration- and cause-specific time effects, using both
completed and administratively censored responses for estimation.
The fitted distribution yields probabilities of correct completion,
incorrect completion, and remaining unfinished, together with
expected token consumption, across budgets within the observation
window. We study CRATER using language-model responses to MATH
questions \citep{hendrycks2021math}. % <!--ref:hendrycks2021math--><!--anchor:section:Abstract-->

\begin{figure}[!htbp]
\centering
\includegraphics[width=\textwidth]{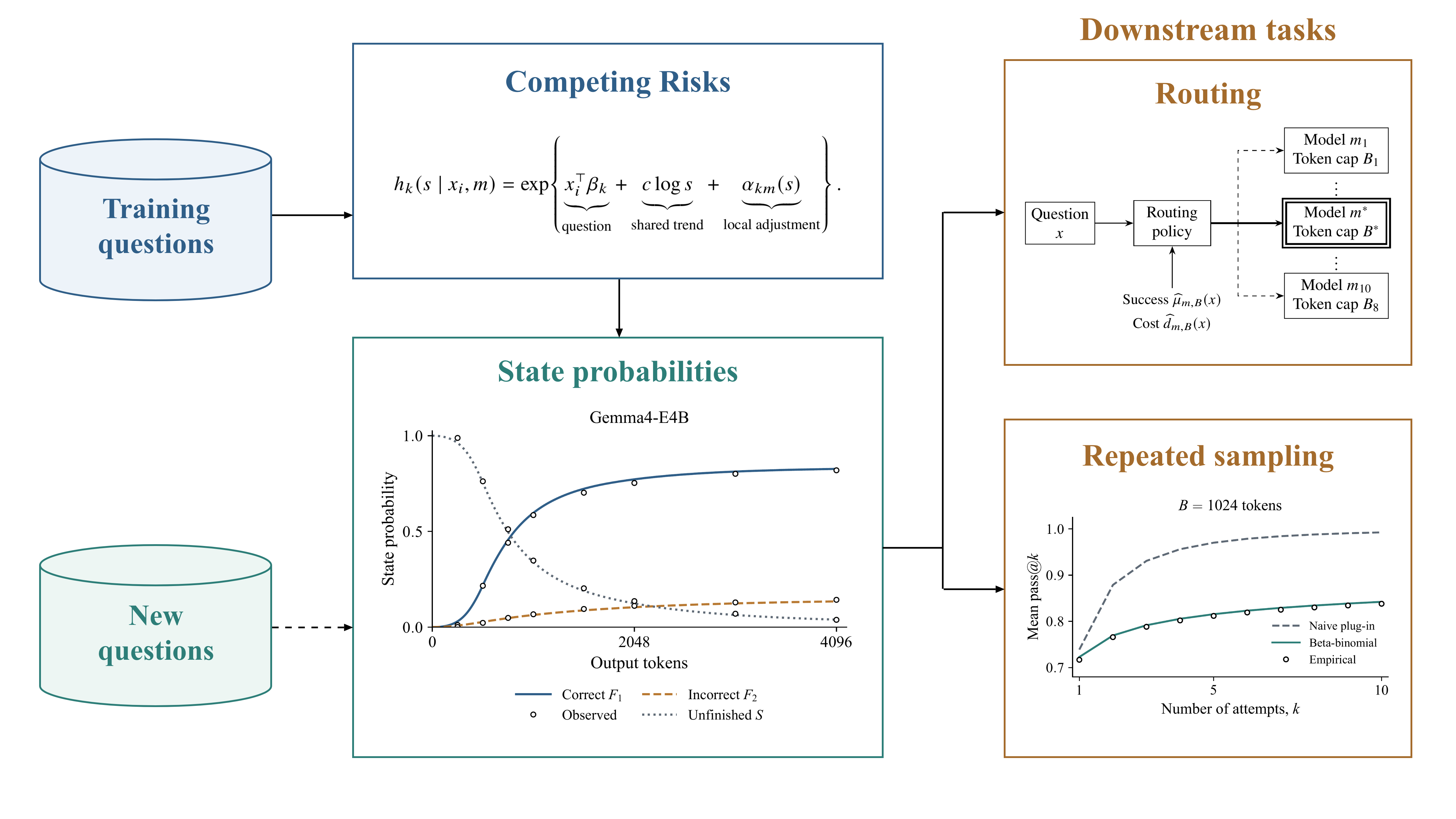}
\caption{Overview of CRATER. Training responses yield a competing-risks model for budget-indexed state probabilities and expected token consumption. Predictions for new questions support model--budget routing and prediction of success under repeated sampling.}
\label{fig:framework}
\end{figure}

We illustrate the value of the fitted distribution in two
downstream tasks.
(i) For routing, predicted success probabilities and expected consumption guide the choice of a configuration and token budget before generation. Under the primary hypothetical prices, the CRATER-based routers achieve lower mean cost than nearest-neighbor predictors using the corresponding question representations, with similar observed correct-completion rates; the magnitude of savings depends on the price assumptions. (ii) For repeated sampling, the model supplies budget-specific probabilities for predicting whether at least one of $k$ attempts succeeds, the quantity summarized by pass@$k$ \citep{chen2021code}. % <!--ref:chen2021code--><!--anchor:section:2.1-->
We connect these predictions to a beta-binomial model, following \citet{kazdan2025passk}, to adjust the predicted mean and account for residual variation in success probabilities when calculating repeated-sampling success. % <!--ref:kazdan2025passk--><!--anchor:section:4.1-->
This adjustment improves agreement with average empirical pass@$k$ in the reported comparisons, which cover up to ten attempts per question. Both applications use the same underlying description of completion time and correctness. Figure~\ref{fig:framework} summarizes the framework.
Our contributions are threefold.
\begin{enumerate}
\item We formulate budget-constrained language-model reasoning as a competing-risks problem, jointly describing completion length and correctness while retaining information from responses right-censored at the generation limit.
\item We develop CRATER, a conditional working model combining
question-embedding features with configuration- and cause-specific
time adjustments. We estimate the model by penalized likelihood,
using a slope-fusion penalty to control the complexity of the
time adjustments.

\item We apply CRATER to model-budget routing and repeated-sampling
prediction using MATH responses. Both applications obtain predictions
at any specified token budget within the observation window from
the same fitted distribution, without refitting the completion
model for each budget.
\end{enumerate}

The Python implementation of CRATER and scripts for reproducing the reported analyses are available at \url{https://anonymous.4open.science/r/llm-competing-risks-4463}.

The rest of the paper is organized as follows. Section~\ref{sec:data} defines the observations and introduces the data. Section~\ref{sec:methods} presents the CRATER working model and estimation procedure. Section~\ref{sec:results} reports parameter estimates, model fit, and component ablations. Section~\ref{sec:downstream} develops the routing and repeated-sampling applications, followed by a discussion of limitations and further directions in Section \ref{sec:discuss}.

\FloatBarrier
% ===== END introduction.tex =====

% ===== BEGIN data.tex =====
\section{Problem setup and dataset}
\label{sec:data}

\subsection{Budgeted reasoning tasks}
\label{sec:reasoning-data}

A budgeted reasoning task asks a language model to answer a question using a limited number of output tokens. We call the output from one generation attempt ``a response sample", or simply ``a sample". It may contain intermediate reasoning as well as a proposed final answer. Generation stops when the model finishes or its token allowance is exhausted. A completed response can be correct or incorrect under a task-specific assessment rule, while an interrupted response may have no final answer. The outcome therefore depends on both the time required to complete the response and the correctness of that completion.

A generation experiment combines a collection of questions, a set of model configurations, a stopping rule, and an assessment procedure. A configuration specifies the model and the conditions under which it generates responses, including prompting and decoding settings. Repeated sampling for the same question under the same configuration can produce different texts, completion times, and answers. Output-token count provides an observable measure of generation length; its relationship to computation, latency or price also depends on the configuration and serving environment.

Let $i$ index questions, $m$ model configurations and $r$ repeated response samples. The question text $q_i$ includes the problem statement and any task context available before generation. For the response sample $(i,m,r)$, let $T_{imr}$ be the natural completion time measured in output tokens, and let $J_{imr}=1$ or $2$ indicate whether the completed response is correct or incorrect according to the assessment rule.

With a prespecified token cap $C_{imr}$, we observe the recorded length $\widetilde T_{imr}=\min(T_{imr},C_{imr})$ and the event indicator

\begin{equation}
\Delta_{imr}=
\begin{cases}
1, & T_{imr}\le C_{imr},\ J_{imr}=1 \quad \text{(correct completion)},\\
2, & T_{imr}\le C_{imr},\ J_{imr}=2 \quad \text{(incorrect completion)},\\
0, & T_{imr}> C_{imr} \quad \text{(right censoring)}.
\end{cases}
\label{eq:event-indicator}
\end{equation}
Thus correctness is recorded only for responses completed before the cap. Response samples reaching the cap remain in the dataset as administratively right-censored records, with natural completion time and eventual correctness unobserved. The strict pre-cap convention reflects the available records, which do not distinguish natural completion exactly at the cap from stopping at the generation limit; Appendix~\ref{app:generation-records} gives the generation and scoring details.

The observed data form the indexed collection

\begin{equation}
\mathcal D_{\mathrm{text}}
=\bigl\{(q_i,m,C_{imr},\widetilde T_{imr},\Delta_{imr})\bigr\}_{(i,m,r)\in\mathcal I},
\label{eq:text-data}
\end{equation}
where $\mathcal I$ indexes the question--configuration--sample combinations present in the experiment. Each record contains the question text, model configuration, token cap, recorded length and event code. The indices $i$ and $r$ organize the observations rather than constitute data fields. The same $q_i$ is shared across configurations and repetitions, while the cap, length and event code refer to an individual response sample. Generated responses, reference answers and assessment records support the construction and checking of the outcomes. They need not be available to a prediction rule applied before generation. Questions provide the natural grouping unit when the aim is to predict performance on other questions, so their repeated observations should remain together in training and evaluation partitions.

For a prospective budget $B$, the prediction targets are the probabilities of correct completion, incorrect completion, and no completion by $B$, conditional on the question and configuration. The probability of a correct completion per response sample combines the chance of finishing with the conditional correctness on finishing. A point estimate of response length or an accuracy calculated only among completed responses describes only part of this problem. Budget-indexed probabilities can inform which configuration to invoke, how many tokens to allocate and how to assess the potential benefit of repeated sampling. Learning these probabilities across questions requires a numerical representation of the question information available before generation.

\subsection{Representing questions with embeddings}
\label{sec:features}

Question text contains information about the task to be solved, but its variable length and symbolic content do not directly form a fixed set of numerical covariates. A text embedding supplies such a representation. Let $e$ be a fixed encoder and define

\begin{equation}
z_i=e(q_i)=\bigl(z_{i1},\ldots,z_{id}\bigr)^\top\in\mathbb R^d.
\label{eq:embedding}
\end{equation}
The encoder maps each question to a vector of the same dimension, allowing statistical models to relate variation in question content to observed completion and correctness. Its coordinates jointly represent aspects of the text; an individual coordinate need not have a direct interpretation as difficulty or a particular reasoning ability.

Representative text encoders include Sentence-BERT, E5 and the BGE family. Sentence-BERT adapts BERT to produce sentence vectors that can be compared through cosine similarity \citep{reimers2019sbert}. % <!--ref:reimers2019sbert--><!--anchor:section:Abstract-->
E5 learns general-purpose text representations through contrastive pretraining on large collections of weakly supervised text pairs \citep{wang2022e5}. % <!--ref:wang2022e5--><!--anchor:section:Abstract-->
BGE-M3 supports multilingual embedding across input lengths ranging from short sentences to longer documents \citep{chen2024m3}. % <!--ref:chen2024m3--><!--anchor:section:Abstract-->
These models provide different implementations of the text-to-vector map, with different training objectives and input capabilities.

Embedding encoders can also be built from generative language models. Qwen3 Embedding uses Qwen3 foundation models as its backbone and combines large-scale pretraining with supervised fine-tuning for text representation \citep{zhang2025embedding}. % <!--ref:zhang2025embedding--><!--anchor:section:Abstract-->
The family provides multilingual embedding models at several parameter scales, including an 8B variant. The encoder and covariate construction used in our analysis are described in Section~\ref{sec:dimension-reduction}.

Since the encoder receives the question before a response is generated, the same $z_i$ is available for all configurations and response samples associated with $q_i$. Replacing $q_i$ by $z_i$ in~\eqref{eq:text-data} gives a numerical data structure consisting of question vectors, configuration identifiers, token caps, recorded lengths and event indicators. A new question can be encoded in the same way before selecting a model or allocating its budget. The input to this representation is the question text, without its generated response, reference solution or observed correctness.

\subsection{Budgeted reasoning on MATH dataset}
\label{sec:math-data}
% ===== BEGIN generated/level5_numbers.tex =====
\newcommand{\LevelFiveATokens}{1478}
\newcommand{\LevelFiveALogprob}{-0.1433}
\newcommand{\LevelFiveBTokens}{3032}
\newcommand{\LevelFiveBLogprob}{-0.1187}
\newcommand{\LevelFiveCTokens}{4096}
\newcommand{\LevelFiveCLogprob}{-0.0484}
\newcommand{\LevelFiveCorrectCount}{3}
\newcommand{\LevelFiveCorrectMinTokens}{1342}
\newcommand{\LevelFiveCorrectMaxTokens}{1590}
\newcommand{\LevelFiveCorrectMinLogprob}{-0.1621}
\newcommand{\LevelFiveCorrectMaxLogprob}{-0.1417}
\newcommand{\LevelFiveIncorrectCount}{1}
\newcommand{\LevelFiveIncorrectMinTokens}{3032}
\newcommand{\LevelFiveIncorrectMaxTokens}{3032}
\newcommand{\LevelFiveIncorrectMinLogprob}{-0.1187}
\newcommand{\LevelFiveIncorrectMaxLogprob}{-0.1187}
\newcommand{\LevelFiveCensoredCount}{6}
\newcommand{\LevelFiveCensoredMinTokens}{4096}
\newcommand{\LevelFiveCensoredMaxTokens}{4096}
\newcommand{\LevelFiveCensoredMinLogprob}{-0.0830}
\newcommand{\LevelFiveCensoredMaxLogprob}{-0.0387}
% ===== END generated/level5_numbers.tex =====

MATH is a benchmark for evaluating mathematical problem solving, containing 12,500 problems from mathematics competitions such as the AMC 10, AMC 12 and AIME \citep{hendrycks2021math}. % <!--ref:hendrycks2021math--><!--anchor:section:3.1 The MATH Dataset-->
The problems are written in natural language and mathematical notation, and their solutions typically involve a sequence of deductions and intermediate calculations.

The questions span seven subjects: prealgebra, algebra, intermediate algebra, precalculus, geometry, number theory, and counting and probability. Each question has a difficulty label from 1 to 5, with 1 denoting the easiest and 5 the hardest problems within a subject. Thus subject and difficulty describe different aspects of the mathematical task.

Each problem comes with a worked reference solution and a final answer. Mathematical expressions are encoded in \LaTeX, and diagrams can be represented by Asymptote code. Final answers, such as numbers, fractions or algebraic expressions, are distinguished from the intermediate reasoning and can be checked after accounting for equivalent mathematical forms.

Responses were generated using nine language-model configurations with 4-bit AWQ weight quantization and one non-quantized configuration, Gemma4-E4B, run in FP16. We use abbreviated model names in the figures and result tables; full checkpoint identifiers and precision settings are listed in Appendix~\ref{app:generation-records}.

Consider the following level-5 number-theory problem from MATH.

\begin{tcolorbox}[
  title={Number theory, Level 5},
  fonttitle=\bfseries,
  coltitle=ExampleBlue,
  colbacktitle=ExampleBlue!12!white,
  colback=ExampleBlue!3!white,
  colframe=ExampleBlue!25!white,
  boxrule=0.4pt,
  arc=2mm,
  outer arc=2mm,
  left=10pt,
  right=10pt,
  top=8pt,
  bottom=8pt,
  before skip=12pt,
  after skip=12pt
]
\textbf{Question.}
Find the least odd prime factor of $2019^8+1$.

\textbf{Answer.} $97$.
\end{tcolorbox}

\paragraph{Different answers to the same question.}
\label{sec:generation-examples}
We use Qwen3-8B-AWQ to illustrate three possible observation types for this question: correct completion, incorrect completion, and right censoring at the token limit. Table~\ref{tab:generation-examples} presents a recorded response for each observation type.

\begin{table}[!ht]
\centering\small
\caption{Three responses to the same level-5 MATH question under Qwen3-8B-AWQ. Excerpts show an opening fragment and the final sentence or recorded ending, separated by an ellipsis. Output markup is reproduced literally, with line breaks normalized and errors retained. Correct and incorrect refer to the recorded answer assessment; censored indicates that no completion is recorded before the token cap.}
\label{tab:generation-examples}
% ===== BEGIN tables/level5_responses.tex =====
\begin{tabular}{@{}p{0.14\linewidth}rp{0.16\linewidth}p{0.53\linewidth}@{}}
\toprule
Response & Tokens & Final answer & Extract from recorded response \\
\midrule
Correct & \LevelFiveATokens & 97 &
{\raggedright\ttfamily We are asked to find the **least odd prime factor** \textnormal{\ldots} So the **smallest odd prime factor** of \$ 2019\textasciicircum{}8 + 1 \$ is: \$\$ \textbackslash{}boxed\{97\} \$\$\par} \\

Incorrect & \LevelFiveBTokens & 353 &
{\raggedright\ttfamily We are asked to find the **least odd prime factor** \textnormal{\ldots} The **least odd prime factor** of \$ 2019\textasciicircum{}8 + 1 \$ is: \$\$ \textbackslash{}boxed\{353\} \$\$\par} \\

Censored & \LevelFiveCTokens & None recorded &
{\raggedright\ttfamily We are asked to find the **least odd prime factor** \textnormal{\ldots} So, \$\$ 2019\textasciicircum{}8 \textbackslash{}equiv 14 \textbackslash{}mod 43 \textbackslash{}Rightarrow N = 2019\textasciicircum{}8 + 1 \textbackslash{}\par} \\
\bottomrule
\end{tabular}
% ===== END tables/level5_responses.tex =====

\end{table}

The correct response uses the order restriction, checks the candidate primes and returns $97$ after \LevelFiveATokens\ tokens. The incorrect response starts with the same restriction but computes the remainder of $6241$ divided by $97$ as $13$ rather than $33$. It consequently rejects $97$ and eventually returns $353$ after \LevelFiveBTokens\ tokens; $353$ is not a divisor of the target number. The censored response tests small primes individually and is still calculating modulo $43$ when its recorded text ends at the cap, without a final answer. This response sample is right-censored at the cap; its natural completion time and eventual correctness remain unobserved.

\begin{figure}[!htb]
\centering\includegraphics[width=\textwidth]{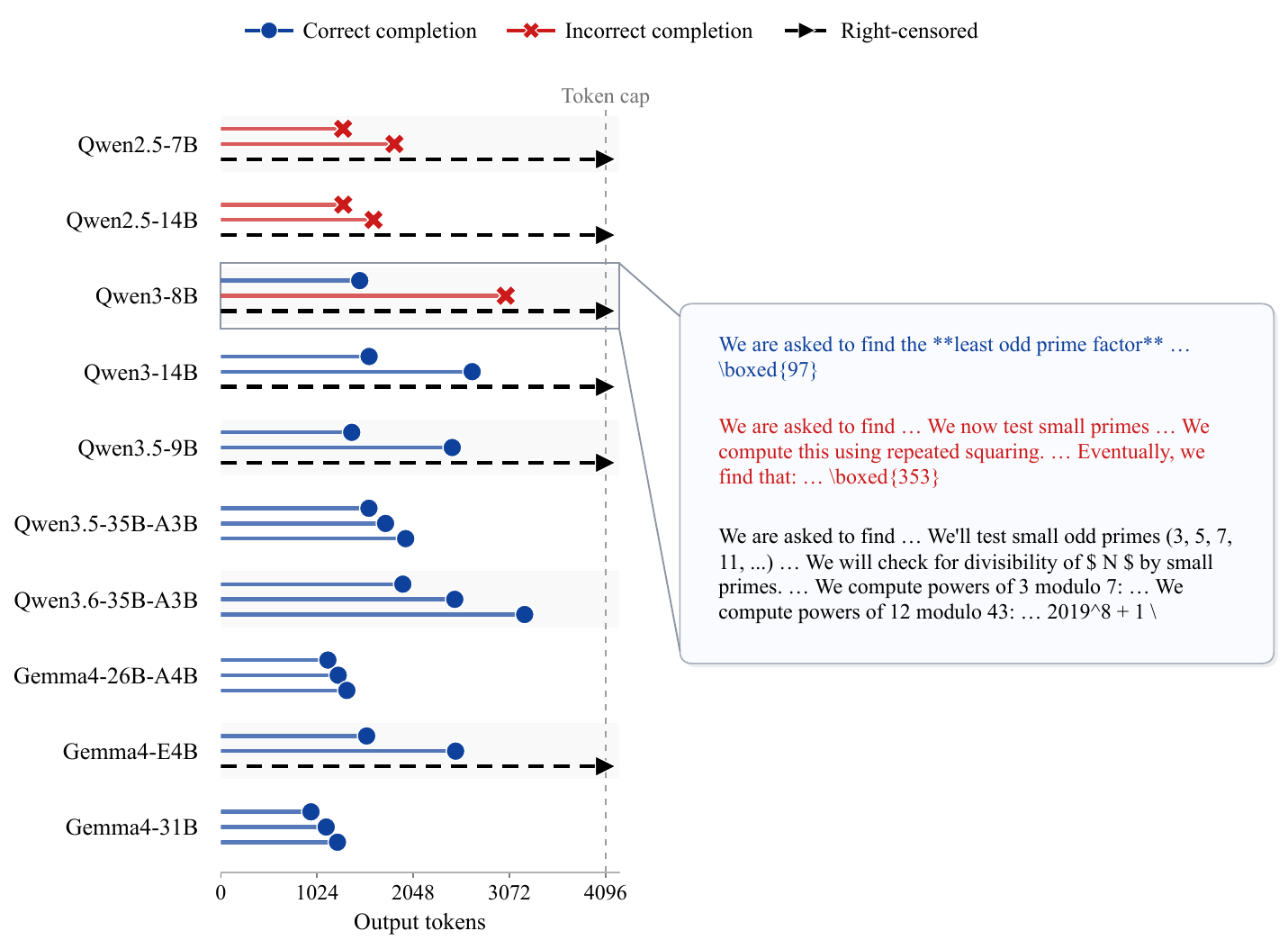}
\caption{Response lengths and outcomes for one level-5 question, with three illustrative response samples per configuration. Circles, crosses and triangles mark correct completion, incorrect completion and right censoring, respectively; the dashed line marks the token cap. Generation settings appear in Appendix~\ref{app:generation-records}.}
\label{fig:level5-example}
\end{figure}

\paragraph{Differences across model configurations.}
Figure~\ref{fig:level5-example} extends the same question to ten model configurations, showing three selected responses from each. Each horizontal line represents one response sample, with its length giving the recorded output-token count and its endpoint marking correct completion, incorrect completion or right censoring. The Qwen3-8B lines show the same three responses as Table~\ref{tab:generation-examples}. The selected Gemma4-31B responses finish correctly at shorter lengths, whereas the selected Qwen2.5 responses either finish incorrectly or reach the cap. The examples illustrate variation in both completion length and correctness across configurations and repeated response samples; a longer response need not end in a correct answer.

For a new question, the response and its outcome are unknown before generation. The statistical task is to use the question and configuration to predict the probabilities of correct completion, incorrect completion and remaining unfinished at each budget. Section~\ref{sec:methods} develops a joint model for those targets.
% ===== END data.tex =====

% ===== BEGIN methods.tex =====
\section{CRATER Framework}
\label{sec:methods}

CRATER jointly models completion length and correctness using output-token count as the time scale. Correct and incorrect completion are treated as competing events, while responses unfinished at the generation limit are administratively right-censored. The working model combines question features with configuration- and cause-specific time effects, estimated by penalized likelihood.

\subsection{Competing risks and token-grid observations}
\label{sec:competing-risks}

Let $X_n$ collect the question information and model configuration for record $n$. With a common generation limit $C$, the observed data are

\begin{equation}
\mathcal D
=\bigl\{(X_n,\widetilde T_n,\Delta_n)\bigr\}_{n=1}^N,
\qquad \widetilde T_n=\min(T_n,C),\quad \Delta_n\in\{0,1,2\}.
\label{eq:data-original}
\end{equation}
Here $\Delta_n=1$ denotes the recorded correct completion, $\Delta_n=2$ the recorded incorrect completion, and $\Delta_n=0$ the administrative right censoring at the generation limit. The numerical representation of the question information is specified in Section~\ref{sec:model-specification}.

Suppose first that completion time $T$ is continuous. The standard modeling framework for continuous-time competing-risks data, with correct and incorrect completion as the two event types and the generation limit providing administrative right censoring \citep{putter2007}, can be naturally adopted. % <!--ref:putter2007--><!--anchor:section:3.2-->
Write $\bar h_1$ and $\bar h_2$ as cause-specific hazards on the original time scale for the two events:

\begin{equation}
\bar h_k(t\mid X)
=\lim_{\varepsilon\downarrow0}
\frac{\Pr(t\leq T<t+\varepsilon,J=k\mid T\geq t,X)}{\varepsilon},
\qquad k=1,2.
\label{eq:hazard-definition}
\end{equation}
Both hazards condition on the response still being unfinished. The survival function of $T$ is

\[
\bar S(t\mid X)
=\exp\left\{-\int_0^t[\bar h_1(u\mid X)+\bar h_2(u\mid X)]\,du\right\}.
\]
\begin{remark}[Completion-conditioned accuracy]
\label{rem:completion-accuracy}
Let $\bar F_k(t\mid X)=\Pr(T\leq t,J=k\mid X)$ denote the probability of type-$k$ completion by time $t$. Accuracy among completed responses at budget $t$ is
\begin{equation}
A_{\mathrm{comp}}(t\mid X)
=\frac{\bar F_1(t\mid X)}{\bar F_1(t\mid X)+\bar F_2(t\mid X)}
=\frac{\bar F_1(t\mid X)}{1-\bar S(t\mid X)},
\label{eq:completion-accuracy}
\end{equation}
provided the denominator is positive. Thus success within a given budget depends on both the completion probability and the completion-conditioned accuracy.
\end{remark}

A type-$k$ completion contributes density $\bar h_k(\widetilde T_n\mid X_n)\bar S(\widetilde T_n\mid X_n)$ to the likelihood, while a censored record contributes $\bar S(C\mid X_n)$. For independent records, the conditional likelihood is therefore
\begin{equation}
L_{\mathrm{cont}}
=\prod_{n=1}^N\left\{
\prod_{k=1}^2
\bigl[\bar h_k(\widetilde T_n\mid X_n)\bar S(\widetilde T_n\mid X_n)\bigr]^{\ind(\Delta_n=k)}
\bar S(C\mid X_n)^{\ind(\Delta_n=0)}
\right\}.
\label{eq:continuous-likelihood}
\end{equation}
Each record contributes exactly one survival factor. Only a recorded completion contributes a cause-specific hazard factor.

Although completion lengths are recorded as integer token counts, $\widetilde T_n\in\{1,\ldots,C\}$, we use a continuous-time competing-risks model as a working description of completion over the token budget. We evaluate its event densities at the recorded token lengths, retaining ties, and use survival contributions for records coded as administratively censored. This gives the working likelihood

\begin{align}
L_{\mathrm{tok}}
=\prod_{n=1}^N\Bigg\{&
\bar h_1(\widetilde T_n\mid X_n)^{\ind(\Delta_n=1)}
\bar h_2(\widetilde T_n\mid X_n)^{\ind(\Delta_n=2)}\notag\\
&\times\exp\left[-\int_0^{\widetilde T_n}
\{\bar h_1(u\mid X_n)+\bar h_2(u\mid X_n)\}\,du\right]
\Bigg\}.
\label{eq:token-likelihood}
\end{align}

For censored records, $\widetilde T_n=C$, so the exponential term is the survival contribution to the cap. We use this likelihood as a working estimation criterion for the recorded token counts. Products over records use working independence; all configurations and response samples of a question remain together when splitting or resampling the data.

When fitting the model, we normalize the token length by the generation limit. Specifically, define
\begin{equation}
\mathcal D'
=\bigl\{(X_n,\widetilde T'_n,\Delta_n)\bigr\}_{n=1}^N,
\qquad \widetilde T'_n=\widetilde T_n/C,\quad s_n=\widetilde T'_n.
\label{eq:data-normalized}
\end{equation}
The observation window then ends at 1. Note that compared with the token counts, the normalized length becomes closer to continuous time. Then, all subsequent model specifications and likelihood calculations use normalized time, with cause-specific hazards denoted by $h_k(s\mid X)$. We write $L_{\mathrm{work}}$ for the working likelihood in~\eqref{eq:token-likelihood} expressed on this scale.

On this scale, survival and the cumulative incidence of type-$k$ completion satisfy

\begin{align}
S(s\mid X)
&=\exp\left\{-\int_0^s[h_1(u\mid X)+h_2(u\mid X)]\,du\right\},
\label{eq:survival-hazards}\\
F_k(s\mid X)
&=\int_0^s S(u\mid X)h_k(u\mid X)\,du,
\qquad k=1,2.
\label{eq:cif}
\end{align}
Thus $F_1$, $F_2$ and $S$ give the probabilities of correct completion, incorrect completion and remaining unfinished, with $F_1+F_2+S=1$. A token budget $B$ corresponds to $s=B/C$. Only the finite observation window is modeled.

\subsection{Model specification}
\label{sec:model-specification}
\label{sec:dimension-reduction}

In CRATER, each cause-specific hazard combines question information, a shared time trend and a local time adjustment specific to the model configuration and completion type. With $x_i$ denoting the features extracted from question $q_i$, we decompose the normalized hazards for each completion type $k$ as

\begin{equation}
h_k(s\mid x_i,m)
=\exp\left\{
\underbrace{x_i^\top\beta_k}_{\text{question information}}
+\underbrace{c\log s}_{\text{shared time trend}}
+\underbrace{\alpha_{km}(s)}_{\text{local time adjustment}}
\right\},\quad 0<s\leq1.
\label{eq:hazard}
\end{equation}
The first term allows questions to differ, the second describes a common time dependence, and the third allows configurations and completion types to depart from that common trend.

\paragraph{Question information.}
The encoder in Section~\ref{sec:features} supplies information about the problem before a response is generated. For the MATH analysis we use Qwen3-Embedding-8B, which maps each question to $z_i\in\mathbb R^{4096}$ \citep{zhang2025embedding}. % <!--ref:zhang2025embedding--><!--anchor:section:Abstract-->
Directly using all coordinates would introduce a large number of feature coefficients. We therefore seek a low-dimensional representation that retains question information relevant to completion time and correctness.

An exploratory comparison of dimension-reduction methods motivates using linear discriminant analysis (LDA) to obtain a compact representation guided by subject and difficulty labels; Appendix~\ref{app:representation-pilot} reports the comparison and its interpretation.

We apply LDA directly to the Qwen embeddings, training separate projections using the difficulty, and subject labels of the training questions. The five difficulty levels yield four discriminant directions, and the seven subjects yield six. Denote the corresponding projection matrices by $V_{\mathrm{level}}\in\mathbb R^{4096\times4}$ and $V_{\mathrm{subject}}\in\mathbb R^{4096\times6}$. Concatenating these directions gives the ten-dimensional question representation

\begin{equation}
W=[V_{\mathrm{level}}\mid V_{\mathrm{subject}}],\qquad
\phi(q_i)=W^\top z_i\in\mathbb R^{10}.
\label{eq:lda}
\end{equation}
We obtain $x_i$ by centering and scaling each coordinate of $\phi(q_i)$ using training means and standard deviations. Once the projection and scaling parameters are fitted, the transformation requires only a new question's text. The resulting covariates enter the log hazard through $x_i^\top\beta_k$, with $\beta_k$ shared across configurations for each completion type $k$.

\begin{remark}[Question information without dimension reduction]
\label{rem:original-embeddings}
As an alternative to dimension reduction, we can model
the question effect using the standardized original
embedding coordinates.
Their high dimensionality and potential collinearity can make coefficient estimation unstable and increase the risk of overfitting. For this extension, we add a ridge penalty $\frac{\rho}{2}\sum_{k=1}^2\|\beta_k\|_2^2$, with $\rho>0$, to regularize the question coefficients while retaining the same time-effect specification. Section~\ref{sec:routing} examines this extension in routing. Appendix~\ref{app:embedding-reparameterization} describes an exact orthogonal reparameterization used for computation.
\end{remark}

\paragraph{Shared time trend.}
We use $\log s$ to provide a parsimonious shared time trend, with the term $c\log s$ corresponding to a multiplicative factor $s^c$ in the hazard. Together with the piecewise-linear adjustment $\alpha_{km}(s)$, this form facilitates computation of the cumulative hazard and likelihood. The adjustment accommodates configuration- and cause-specific departures from the shared trend. Sensitivity to the choice of $c$ is examined in Appendix~\ref{app:power-sensitivity}.

\paragraph{Local time adjustment.}
To allow departures from the shared trend, we use a continuous piecewise-linear function for local time adjustment,

\begin{equation}
\alpha_{km}(s)=a_{km}+b_{km}s+
\sum_{j=1}^{K}\gamma_{kmj}(s-\kappa_j)_+,
\qquad (v)_+=\max(v,0).
\label{eq:spline}
\end{equation}
The intercept, initial slope, and subsequent slope changes are specific to configuration $m$ and completion type $k$. The main fit uses $K=18$ candidate knots, placed more densely near the beginning of the observation window; their locations are listed in Appendix~\ref{app:development-design}. At $\kappa_j$, $\gamma_{kmj}$ is the change between the slopes of adjacent intervals. Penalizing $\sum_j|\gamma_{kmj}|$ therefore applies the fusion component of the fused lasso to the segment slopes, encouraging neighboring intervals to share the same slope \citep{tibshirani2005fused}. % <!--ref:tibshirani2005fused--><!--anchor:page:93-->
Continuity is guaranteed by the hinge basis itself. A nonzero coefficient changes the slope without creating a jump in $\alpha_{km}$ or the hazard. This construction permits local changes while regularizing their extent, and its estimation is described next. The equivalence to a fused-lasso penalty on adjacent segment
slopes is shown in Appendix~\ref{app:math}.

\subsection{Penalized likelihood estimation}
\label{sec:estimation}

For fixed $c$, substitute~\eqref{eq:hazard} into the normalized working likelihood $L_{\mathrm{work}}$. Write $i(n)$ and $m(n)$ for the question and configuration of record $n$. The average negative log likelihood is

\begin{align}
\mathcal L(\theta)
&=\frac1N\sum_{n=1}^N\Bigg[
\sum_{k=1}^2\int_0^{s_n}u^c
\exp\{x_{i(n)}^\top\beta_k+\alpha_{k,m(n)}(u)\}\,du\notag\\
&\hspace{4em}-\sum_{k=1}^2\ind(\Delta_n=k)
\{x_{i(n)}^\top\beta_k+c\log s_n+\alpha_{k,m(n)}(s_n)\}
\Bigg].
\label{eq:loss}
\end{align}
A censored response sample contributes only the integrated hazards, whereas a completed response sample also contributes its event log hazard. Equal numbers of configurations and repetitions per question give equal question weights in this criterion.

The parameter vector $\theta$ collects the feature and spline coefficients. We estimate it by

\begin{equation}
\widehat\theta
=\arg\min_\theta\left\{
\mathcal L(\theta)
+\mathcal P_\lambda(\theta)
\right\},
\label{eq:penalty}
\end{equation}
where $\mathcal P_\lambda$ is the slope-fusion penalty, that is, 

\begin{equation}
\mathcal P_\lambda(\theta)
=\lambda\sum_{k,m}\sum_{j=1}^K|\gamma_{kmj}|.
\label{eq:full-penalty}
\end{equation}
The tuning parameter $\lambda$ controls changes between adjacent segment slopes. In the LDA specification, the question coefficients $\beta_k$, intercepts $a_{km}$ and initial slopes $b_{km}$ are unpenalized. This regularizes the complexity of the time adjustments without selecting question features or shrinking their baseline levels. Appendix~\ref{app:math} discusses the parameterization and optimization.

With the LDA representation fixed, we select $\lambda$ by validation within the training sample. All configurations and response samples for a question remain in the same partition, and feature standardization uses only the training questions in each split. After selecting $\lambda$, we refit the model using all training questions and assess its predictions on the separate evaluation sample.

The integrals in~\eqref{eq:loss} are over time, conditional on the question features. On a spline interval the main hazard is $u\exp(A+Du)$, whose cumulative integral has an elementary form. Prediction additionally requires integrating $S h_k$. Numerical quadrature and a proximal Newton method are used for model training, with the objective, derivatives and probability calculations checked independently. Appendix~\ref{app:numerics} gives the formulas, optimization checks and tolerances.

% ===== END methods.tex =====

% ===== BEGIN results.tex =====
\section{Empirical Performance}
\label{sec:results}

We apply CRATER to the MATH generation data to study how question information and model configuration relate to completion time and correctness. After introducing the evaluation criteria and experimental setup, we report parameter estimates, fitted hazards, and training-sample state probabilities with $c=1$, followed by ablations of question and token effects.

\subsection{Evaluation criteria and experimental setup}
\label{sec:numerical-setup}

\paragraph{Probability evaluation.}
We evaluate predicted correct- and incorrect-completion probabilities using the integrated Brier score (IBS), which is commonly used in survival analysis. For normalized budget $s\in[0,1]$, define $Y_{nk}(s)=\ind(s_n\leq s,\Delta_n=k)$ for completion type $k=1,2$. Let $\mathcal E$ denote the set of generation records being evaluated, with $N_{\mathcal E}=|\mathcal E|$. Each record corresponds to one response sample. For record $n$, let $i(n)$ be its question index and $m(n)$ its model-configuration index. We evaluate the mean of the two cause-specific Brier scores and its integral over the observation window:

\begin{align}
\widehat{\operatorname{BS}}(s)
&=\frac{1}{2N_{\mathcal E}}\sum_{n\in\mathcal E}\sum_{k=1}^2
\left\{\widehat F_k(s\mid x_{i(n)},m(n))-Y_{nk}(s)\right\}^2,
\label{eq:brier}
\end{align}
At each budget, the Brier score averages the squared differences between predicted completion probabilities and the corresponding binary outcomes, giving equal weight to correct and incorrect completion. IBS averages this probability-prediction error over the normalized budget window. Lower values indicate more accurate probability predictions. We apply the same criterion to training, validation and evaluation records.

To approximate the integral on the token grid, let $s_j=j/C$ for every integer budget $j=0,1,\ldots,C$. The trapezoidal approximation is

\begin{equation}
\widehat{\operatorname{IBS}}
\approx \frac{1}{C}\left[
\frac{\widehat{\operatorname{BS}}(0)}{2}
+\sum_{j=1}^{C-1}\widehat{\operatorname{BS}}\!\left(\frac{j}{C}\right)
+\frac{\widehat{\operatorname{BS}}(1)}{2}
\right].
\label{eq:brier-token-grid}
\end{equation}
The factor $1/C$ is the spacing between adjacent normalized budgets. This approximation evaluates the Brier score at every integer token budget and gives equal weight to equal increments of token budget. Since each question has the same numbers of configurations and response samples, the record average also weights questions equally.

The score uses the outcomes of individual response samples. Averaging the ten outcomes before squaring would instead measure error relative to an empirical question-level proportion and would change the criterion. The third state has predicted probability $\widehat S=1-\widehat F_1-\widehat F_2$, so the two cause probabilities determine the full three-state prediction. All observed-state indicators follow the cap convention in Section~\ref{sec:competing-risks}. Evaluation-sample state-probability curves complement the integrated score in Appendix~\ref{app:state-probabilities}.

\paragraph{Data partition and working specification.}
The analysis includes 2787 questions, with ten model configurations and ten response samples per question--configuration pair. We retain the question partition of 1400 training questions and 1387 evaluation questions. All configurations and response samples of a question remain together in every partition and resample. Thus evaluation concerns predictions for other questions under the same collection of configurations.

The main CRATER specification uses the LDA features in Section~\ref{sec:model-specification}, the shared time term $c\log s$, and the continuous piecewise-linear adjustments in~\eqref{eq:spline}. We fix $c=1$ as a convenient working specification. On each spline interval, the hazard then takes the form $s\exp(A+Ds)$, with $A$ and $D$ constant, and admits an elementary cumulative integral. Sensitivity analyses with alternative values of $c$ yield similar validation and evaluation scores (Appendix~\ref{app:power-sensitivity}), indicating that predictive performance is insensitive to this choice within the fitted model family. All subsequent analyses use $c=1$.

The LDA projection is learned from the training questions and held fixed during penalty selection. Within the training sample, we compare candidate values of $\lambda$ by validation on questions excluded from training each outcome model. Feature centering and scaling use the training questions in each split. We select the value with the smallest mean validation IBS in~\eqref{eq:brier-token-grid}, and refit on all training questions. Evaluation questions enter neither the LDA fit nor penalty selection.

We use three subject--difficulty-stratified validation splits, each containing 1120 training questions and 280 validation questions. The initial grid contains 19 penalty values from $10^{-7}$ to $10^{-3}$. A prespecified rule extends a grid boundary by factors of three and ten, for at most two rounds, if the validation minimum occurs there. The time adjustment uses 18 internal knots, held fixed throughout the comparisons.

In the following analysis, evaluation questions are used only to assess the fitted models, not to learn the LDA representation or select penalties. This evaluation set was also examined in earlier experiments.

\subsection{Parameter estimates and model fit}
\label{sec:parameter-estimates}

We report parameter estimates and model fit results under the specification described above.

\paragraph{Parameter estimates.}
Table~\ref{tab:beta} provides the feature coefficients for correct and incorrect completion, together with their bootstrap standard deviations. The first four coordinates are the LDA directions learned from difficulty labels, and the remaining six are learned from subject labels. Each coordinate is standardized using the training data.

A coefficient $\beta_{kj}$ is the change in the log hazard of  cause-$k$ associated with a one-standard-deviation increase in feature $j$, holding the other features and configuration fixed. The associated hazard ratio is $\exp(\beta_{kj})$. Comparing $\beta_{1j}$ and $\beta_{2j}$ shows whether a feature has similar or different associations with the timing of correct and incorrect completion. Their implications for correct-completion probability also depend on both hazards through the survival function.

\begin{table}[!htb]
\centering\small
\caption{Feature coefficient estimates for CRATER with $c=1$. LD1--LD4 are learned from difficulty labels and LD5--LD10 from subject labels. Parentheses give sample standard deviations of coefficient estimates obtained from 200 bootstrap samples, constructed by resampling training questions within subject--difficulty strata and retaining all configurations and response samples of each selected question. The representation and model specification are held fixed in each bootstrap sample.}
\label{tab:beta}
% ===== BEGIN tables/section4_feature_estimates.tex =====
\begingroup
\begin{tabular*}{.88\textwidth}{@{\extracolsep{\fill}}llrr@{}}
\toprule
LDA labels & Feature & \shortstack{Correct completion\\$\widehat\beta_1$} & \shortstack{Incorrect completion\\$\widehat\beta_2$} \\
\midrule
Difficulty & LD1 & $-0.674\;(0.032)$ & $-0.073\;(0.039)$ \\
 & LD2 & $-0.056\;(0.031)$ & $-0.065\;(0.043)$ \\
 & LD3 & $0.048\;(0.028)$ & $-0.020\;(0.048)$ \\
 & LD4 & $0.068\;(0.026)$ & $0.075\;(0.050)$ \\

Subject & LD5 & $-0.340\;(0.026)$ & $-0.158\;(0.036)$ \\
 & LD6 & $0.093\;(0.024)$ & $-0.047\;(0.030)$ \\
 & LD7 & $-0.223\;(0.030)$ & $0.049\;(0.038)$ \\
 & LD8 & $-0.055\;(0.023)$ & $-0.202\;(0.047)$ \\
 & LD9 & $0.272\;(0.030)$ & $-0.043\;(0.045)$ \\
 & LD10 & $-0.143\;(0.038)$ & $-0.153\;(0.051)$ \\
\bottomrule
\end{tabular*}
\endgroup
% ===== END tables/section4_feature_estimates.tex =====

\end{table}

Table~\ref{tab:beta} reports the estimated question-feature coefficients and their bootstrap standard deviations. The estimates differ between correct and incorrect completion, suggesting that question features have different associations with the two cause-specific hazards.

\begin{figure}[!htbp]
\centering
\includegraphics[width=.88\textwidth]{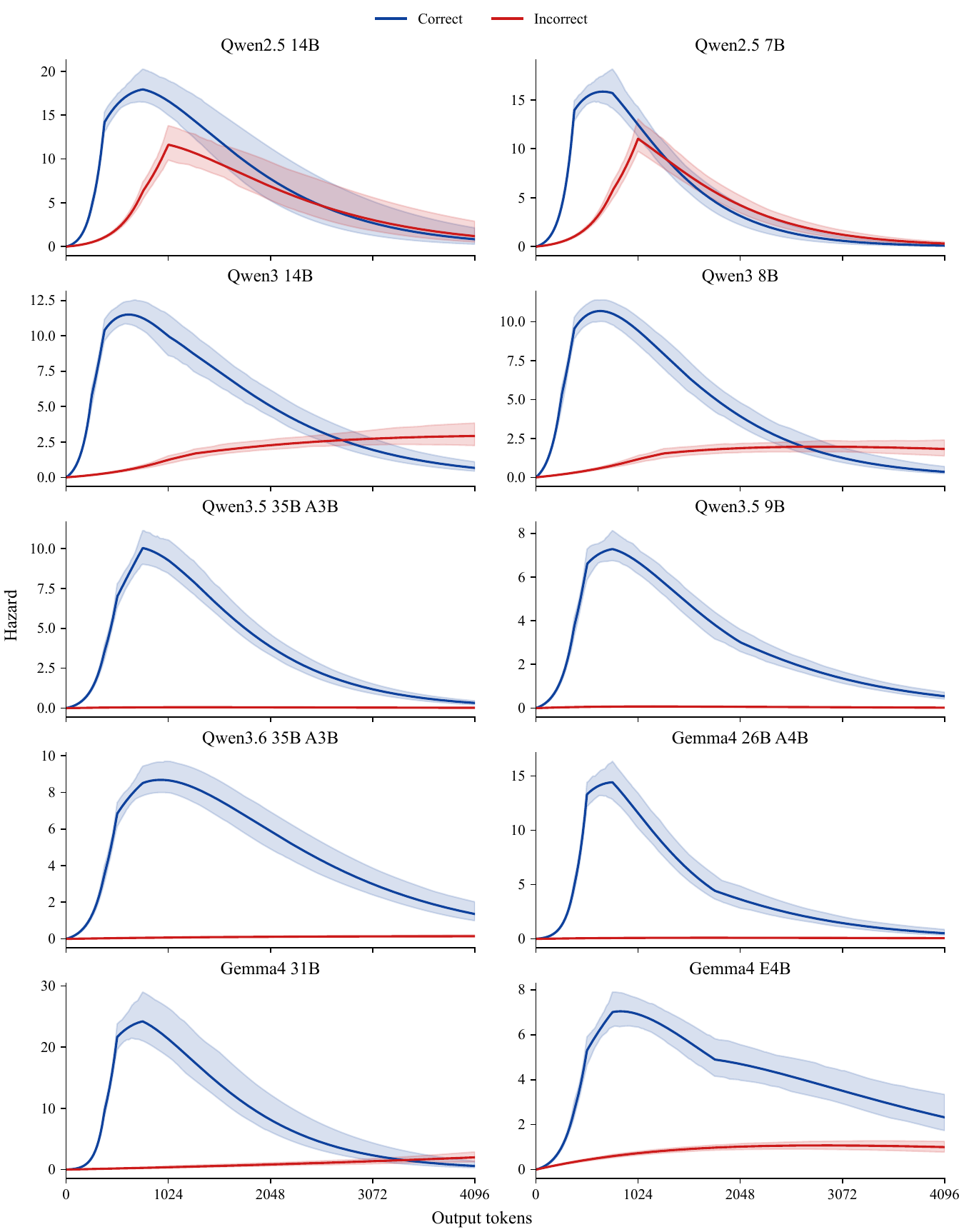}
\caption{Fitted hazards for correct and incorrect completion across model configurations at centered question features ($x=0$). Rates are per unit of normalized token time. Shaded bands show pointwise 95\% percentile intervals from 200 bootstrap samples, resampling questions within subject--difficulty strata and retaining all associated response samples.}
\label{fig:hazards}
\end{figure}

\paragraph{Fitted hazards.}
Figure~\ref{fig:hazards} displays the fitted cause-specific hazards at centered question features, $x=0$. Each hazard describes the instantaneous rate of the corresponding completion among responses still unfinished at that budget. The underlying time adjustments appear in Appendix Figure~\ref{fig:alpha}.

Correct-completion hazards rise early and then decline across all configurations. Beyond the peak, responses still unfinished have a lower rate of correct completion. Correct completions nevertheless continue to accumulate, so the hazard peak alone does not identify an optimal stopping budget.

Incorrect-completion hazards vary across configurations, with similar patterns within the Qwen2.5 and Qwen3 families. Both Qwen2.5 configurations show a later peak for incorrect than for correct completion, whereas both Qwen3 configurations maintain elevated incorrect-completion hazards at longer budgets. For these configurations and Gemma4-31B, the incorrect-completion hazard eventually exceeds the correct-completion hazard. Qwen3.5, Qwen3.6 and Gemma4-26B retain lower incorrect-completion hazards throughout the plotted positive budgets. 

The evaluation results show different patterns of unsuccessful
responses across configurations with lower and higher
correct-completion rates (Appendix Table~\ref{tab:data}).
The weaker configurations, Qwen2.5 and Qwen3, generally
produce shorter responses, and their unsuccessful responses
predominantly end in incorrect answers.
The stronger configurations, Qwen3.5, Qwen3.6 and
Gemma4-26B, generally produce longer responses, and their
unsuccessful responses more often exhaust the token budget
without completing.
Gemma4-31B is a notable exception: it combines the highest
correct-completion rate with short responses and few cap hits.
These patterns distinguish configurations whose unsuccessful
responses mostly end in incorrect answers from those whose
unsuccessful responses mostly exhaust the token budget.
An overall correct-completion rate does not capture
this distinction.

\begin{figure}[!htbp]
\centering
\includegraphics[width=\textwidth]{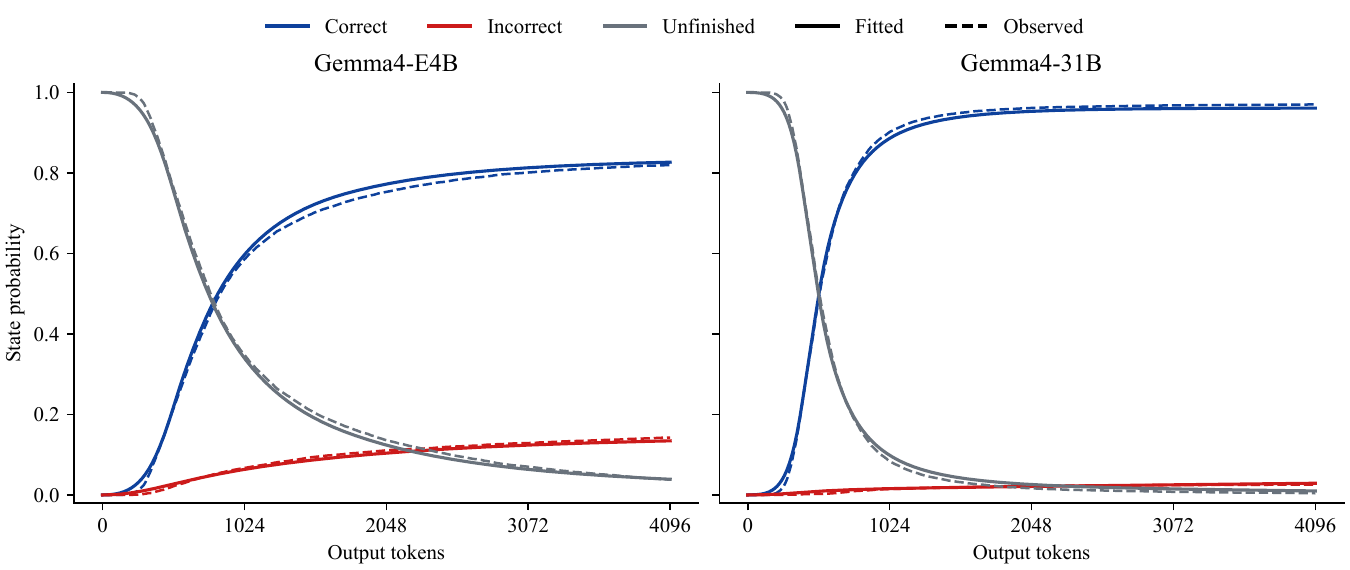}
\caption{Overall fitted and observed state probabilities for Gemma4-E4B (left) and Gemma4-31B (right) on 1400 training questions. Solid curves average fitted probabilities over questions; dashed curves show observed proportions across all associated response samples.}
\label{fig:states-training}
\end{figure}

\paragraph{Fitted and observed state probabilities.}
Figure~\ref{fig:states-training} compares fitted state probabilities with observed proportions across all 1400 training questions for Gemma4-E4B and Gemma4-31B, using the same fitted model as Figure~\ref{fig:hazards}. These configurations illustrate lower and higher overall correct-completion rates, respectively. Solid curves average fitted probabilities over questions; dashed curves give the fractions of response samples completed correctly, completed incorrectly or still unfinished at each token budget.

The fitted curves closely follow the overall completion profiles of both configurations. Correct completions accumulate earlier and reach a higher final proportion for Gemma4-31B, whereas Gemma4-E4B has larger incorrect and unfinished proportions. These pooled comparisons describe in-sample fit; difficulty-specific discrepancies and evaluation-sample diagnostics are reported in Appendices~\ref{app:training-states} and~\ref{app:state-probabilities}.

\subsection{Ablation of question and token effects}
\label{sec:ablation}

We examine the question and token effects in model (\ref{eq:hazard}) through a complete $2\times2$ ablation, retaining or removing each component. The first ablation removes $x_i^\top\beta_k$, retaining the shared $\log s$ trend and local adjustment. The second removes both $\log s$ and the time-varying part of $\alpha_{km}(s)$, giving
\begin{equation}
h_k(s\mid x_i,m)=\exp\{x_i^\top\beta_k+a_{km}\}.
\label{eq:ablation-no-token}
\end{equation}
This model retains question effects and a separate baseline level for each configuration and completion type, but its hazards are constant over token budget. Its cumulative completion probabilities still increase with the budget through the survival function. The final ablation removes both components, leaving the intercept-only model $h_k(s\mid m)=\exp(a_{km})$. It assigns the same hazards to all questions within a configuration, while retaining separate baseline levels for correct and incorrect completion.

All four models are trained on the same 1400 training questions and scored on both the training and evaluation samples. 
Each ablation retains the settings of the main specification, with the remaining coefficients reestimated.

\begin{table}[!htbp]
\centering\small
\caption{Training and evaluation integrated Brier scores (IBS) for ablations of question and token effects. Values in parentheses are bootstrap standard deviations from 4000 subject--difficulty-stratified question resamples within each sample, keeping all configurations and response samples and holding the fitted predictors fixed. All four models retain configuration- and cause-specific intercepts; the final row contains no question or token effects. Lower IBS indicates better probability prediction.}
\label{tab:brier}
% ===== BEGIN tables/section4_ablation.tex =====
\begin{tabular}{lcc}
\toprule
Model specification & Training IBS & Evaluation IBS \\
\midrule
CRATER (full model) & \textbf{0.07382 (0.00201)} & \textbf{0.09103 (0.00237)} \\
Without question effects & 0.09085 (0.00221) & 0.09727 (0.00234) \\
Without token effects & 0.07824 (0.00186) & 0.09387 (0.00222) \\
Intercept-only baseline & 0.09370 (0.00203) & 0.09959 (0.00217) \\
\bottomrule
\end{tabular}
% ===== END tables/section4_ablation.tex =====

\end{table}

Table \ref{tab:brier} shows that removing either question effects or token effects increases IBS on both the training and evaluation sets, with larger increases when question effects are removed. On this evaluation sample, CRATER reduces IBS by 6.4\% relative to the model without question effects and by 3.0\% relative to the model without token effects. Both question information and token-dependent hazards therefore contribute to the observed probability fit and predictive performance. Relative to the intercept-only baseline, either component improves prediction, with question effects yielding the larger reduction in evaluation IBS.

The token-effect ablation removes both the shared trend and the local time adjustment; sensitivity to the shared time coefficient is examined separately in Appendix~\ref{app:power-sensitivity}.

\FloatBarrier
% ===== END results.tex =====

% ===== BEGIN downstream.tex =====
\section{Downstream tasks}
\label{sec:downstream}

We consider two downstream applications of CRATER, routing questions to model configurations and token budgets, and predicting success probabilities under repeated sampling.

% ===== BEGIN routing.tex =====
\subsection{Routing before generation}
\label{sec:routing}

Language-model routing selects a configuration for an incoming question before generating its response, balancing expected answer quality against inference cost. Existing approaches use question information to predict which configuration will perform well: RouteLLM learns from model preferences, RouterBench evaluates nearest-neighbor and neural predictors, and CARROT predicts both performance and cost \citep{ong2024route,hu2024routerbench,somerstep2025carrot}. % <!--ref:ong2024route--><!--anchor:section:Abstract--><!--ref:hu2024routerbench--><!--anchor:section:5.1--><!--ref:somerstep2025carrot--><!--anchor:section:4.2-->
Budget selection adds a second decision. R2-Router jointly chooses a configuration and an output-length budget using estimated quality--cost curves \citep{xue2026r2}. % <!--ref:xue2026r2--><!--anchor:section:Abstract-->
Here we compare nearest-neighbor routing with CRATER-based routing under two question representations, using the same configuration and budget choices within each comparison.

Let $a=(m,B)\in\mathcal A$ denote an action that assigns configuration $m$ and token cap $B$ to a question. With a fixed cap, this reduces to choosing a configuration alone. For question features $x$, let $\mu_a(x)=\mu_{m,B}(x)$ be the probability of correct completion by $B$, and let $d_a(x)=d_{m,B}(x)$ be the expected generation cost. A routing policy specifies the probability $\pi_a(x)$ of selecting each action, with $\pi_a(x)\geq0$ and $\sum_{a\in\mathcal A}\pi_a(x)=1$. The quality--cost objective is

\begin{equation}
\max_{\pi}\ \E_X\!\left[\sum_{a\in\mathcal A}\pi_a(X)\mu_a(X)\right],
\qquad
\E_X\!\left[\sum_{a\in\mathcal A}\pi_a(X)d_a(X)\right]\leq b,
\label{eq:routing-objective}
\end{equation}
where $b$ is an average cost allowance. The sums average over the policy's action probabilities, and $\E_X$ averages over the population of incoming questions. For a fixed cost multiplier $\eta\geq0$, maximizing the Lagrangian pointwise gives a deterministic candidate rule

\begin{equation}
g_\eta(x)\in\arg\max_{a\in\mathcal A}
\{\mu_a(x)-\eta d_a(x)\}.
\label{eq:routing-standard}
\end{equation}
Here $g_\eta(x)$ returns one model--budget pair, whereas $\pi_a(x)$ describes a possibly randomized policy. Larger $\eta$ gives greater weight to cost. This formulation separates prediction of quality and cost from the allocation decision \citep{somerstep2025carrot}. % <!--ref:somerstep2025carrot--><!--anchor:section:4.2-->

In our setting, quality and cost also depend on when generation is stopped. A response may complete correctly, complete incorrectly, or remain unfinished at its cap. Separate predictions at selected budgets need not correspond to a common completion-time distribution. CRATER supplies that connection through $F_1$, $F_2$ and $S$, allowing completed and right-censored records to inform the entire budget range. We evaluate smaller budgets using the recorded responses, assuming that changing the cap leaves the distribution of generated prefixes unchanged under fixed prompts and decoding settings. Success requires a recorded correct completion by the assigned budget; unfinished prefixes are not rescored.

For an action $(m,B)$, let $p_m$ be configuration $m$'s price per output token and $B\leq C$ its allowed budget. The expected number of tokens actually used is $L_m(B\mid x)=\E\{\min(T,B)\mid x,m\}$. Thus CRATER gives

\begin{align}
\widehat\mu_{m,B}(x)&=\widehat F_1(B/C\mid x,m),\notag\\
\widehat d_{m,B}(x)&=p_m\widehat L_m(B\mid x),
\label{eq:cost}
\end{align}
where $\widehat L_m(B\mid x)=C\int_0^{B/C}\widehat S(s\mid x,m)\,ds$ is the predicted number of output tokens used before completion or the cap $B$. Substituting these estimates into~\eqref{eq:routing-standard} gives

\begin{equation}
\widehat g_\eta(x)\in\arg\max_{(m,B)\in\mathcal A}
\left\{\widehat F_1(B/C\mid x,m)-\eta p_m\widehat L_m(B\mid x)\right\}.
\label{eq:priced-policy}
\end{equation}
Each value of $\eta$ defines a candidate rule that selects one action for each question; the final policy is chosen by validation as described below.

\paragraph{Routing methods and question representations.}
We compare $k$-nearest-neighbor ($k$NN) and CRATER predictors in two pairs: one uses the original 4096-dimensional Qwen question embeddings, and the other uses their ten-dimensional LDA representation. A strong reference and a question-independent policy provide common benchmarks.
\begin{enumerate}
\item \emph{Strong reference.} Every question is assigned to Gemma4-31B at the maximum budget $C$, selected as the reference by its training correct-completion rate.
\item \emph{Zero Router.} Adapting the question-independent baseline of \citet{hu2024routerbench}, we estimate the correct, incorrect and unfinished proportions and mean cost for each configuration--budget action on the validation questions. A linear program selects fixed probabilities for these actions to minimize mean cost subject to the quality constraints below. Each new question receives one action sampled from this same distribution, without using its content. % <!--ref:hu2024routerbench--><!--anchor:section:3.2-->
\item \emph{$k$NN.} Nearest-neighbor prediction is used for routing in RouterBench and CARROT \citep{hu2024routerbench,somerstep2025carrot}. We adapt this approach to configuration--budget actions by averaging correct-by-budget indicators and token consumption over neighboring training questions, converting predicted consumption to cost using the configuration's token price. We evaluate two versions: one uses the original embeddings with cosine distance, and the other uses the ten-dimensional LDA features with Euclidean distance after coordinate standardization. Each version is compared with CRATER using the corresponding question representation. The predicted success probabilities and costs enter the action-selection rule underlying~\eqref{eq:priced-policy}. % <!--ref:hu2024routerbench--><!--anchor:section:5.1--><!--ref:somerstep2025carrot--><!--anchor:section:4.2-->
\item \emph{CRATER.} We estimate budget-specific success probabilities and expected costs from the fitted competing-risks distribution via~\eqref{eq:cost} and use these estimates in~\eqref{eq:priced-policy}. We consider both the LDA specification in Section~\ref{sec:results} and the original-embedding specification with ridge-penalized question coefficients in Remark~\ref{rem:original-embeddings}. Training details are provided in Appendix~\ref{app:pricedrouting}.
\end{enumerate}
Within each representation, both predictors choose a configuration and a budget from the same action set. Their preprocessing and regularization follow their respective training procedures.

\paragraph{Experimental setup.}
The candidate actions combine ten configurations with eight caps: 256, 512, 768, 1024, 1536, 2048, 3072, and 4096 tokens. All four question-dependent predictors and Zero Router use these 80 actions. We assign hypothetical prices based on active parameter count and training performance (Table~\ref{tab:routing-prices}). Let $A_m$ be the active parameter count in billions, using the total count for dense models, and let $r_m$ rank training full-cap correct-completion rates from lowest to highest. We set $q_m=(r_m-1)/9$ and $p_m=A_m(1+q_m)$. The capability multiplier ranges from one to two. A response sample assigned cap $B$ incurs cost $p_m\min(\widetilde T,B)$; these prices exclude input processing and serving overhead.

\begin{table}[!htbp]
\centering\small
\setlength{\tabcolsep}{5pt}
\caption{Hypothetical per-output-token prices, in arbitrary cost units. Correct-completion rates average ten response samples on each of the 1400 training questions at the full cap. Prices follow $p_m=A_m(1+q_m)$.}
\label{tab:routing-prices}
% ===== BEGIN tables/capability_routing_prices.tex =====
\begin{tabular}{lrrr}
\toprule
Configuration & Active parameters (B) & Training correct (\%) & Price $p_m$ \\
\midrule
Qwen2.5-7B & 7 & 79.71 & 7.00 \\
Qwen2.5-14B & 14 & 81.41 & 15.56 \\
Gemma4-E4B & 4 & 81.93 & 4.89 \\
Qwen3-8B & 8 & 85.55 & 10.67 \\
Qwen3.5-9B & 9 & 87.34 & 13.00 \\
Qwen3-14B & 14 & 87.40 & 21.78 \\
Qwen3.5-A3B & 3 & 90.80 & 5.00 \\
Gemma4-A4B & 4 & 93.67 & 7.11 \\
Qwen3.6-A3B & 3 & 94.04 & 5.67 \\
Gemma4-31B & 31 & 97.01 & 62.00 \\
\bottomrule
\end{tabular}
% ===== END tables/capability_routing_prices.tex =====

\end{table}

All policy choices use validation questions within the training sample. For the experiment, we minimize validation cost subject to quality requirements, rather than fix a cost allowance in~\eqref{eq:routing-objective}. Relative to the strong reference, we allow at most a one-percentage-point decrease in correct completion, a half-point increase in incorrect completion, and a one-point increase in unfinished responses. We construct candidate rules from a grid of $\eta$ values and include the strong-reference rule. A linear program chooses their mixture probabilities to minimize observed validation cost under these constraints. For each new question, the resulting policy samples a rule with the selected probabilities and applies it to the question's features. The rules and mixture probabilities are held fixed for evaluation. Appendix~\ref{app:pricedrouting} gives the grid and selection program. Every evaluation question receives one generation action; offline replay averages over its ten recorded response samples and the policy's randomization.

\paragraph{Numerical results.}
Table~\ref{tab:routing} reports results within each representation. Zero Router attains \RPLPAccuracy\% correct completion and reduces cost by \RPLPSaving\% relative to the strong reference. With original embeddings, $k$NN and CRATER with ridge attain \RPEmbeddingKnnAccuracy\% and \RPEmbeddingRiskAccuracy\% correct completion, with cost reductions of \RPEmbeddingKnnSaving\% and \RPEmbeddingRiskSaving\% relative to the reference. CRATER's cost is \RPEmbeddingReduction\% below $k$NN's.

\begin{table}[tb]
\centering\small
\setlength{\tabcolsep}{4pt}
\caption{Routing comparisons under the primary validation quality constraints. The first $k$NN--CRATER pair uses ten-dimensional LDA features; the final pair uses original embeddings, with a ridge penalty on CRATER's question coefficients. All adaptive methods use the same 80 actions. Outcomes and costs average ten response samples on each of 1387 evaluation questions, with policy randomization averaged analytically. Mean cost uses the price units in Table~\ref{tab:routing-prices}; parentheses give bootstrap SDs from 4000 paired question resamples within subject--difficulty strata, holding predictors and policies fixed. Negative percentages in the final column indicate cost reductions relative to the strong reference.}
\label{tab:routing}
% ===== BEGIN tables/matched_routing.tex =====
\begin{tabular}{lrrrrr}
\toprule
& \multicolumn{3}{c}{Outcome (\%)} & Mean cost & Cost saving \\
Method & Correct & Incorrect & Unfinished & (SD) & vs. strong (\%) \\
\midrule
Strong reference & 97.30 & 2.38 & 0.32 & 36,878 (444) & --- \\
Zero Router & 96.42 & 2.58 & 1.00 & 29,362 (354) & -20.38 \\
$k$NN (LDA) & 96.27 & 2.25 & 1.48 & 26,928 (444) & -26.98 \\
CRATER (LDA) & 96.40 & 2.11 & 1.49 & \textbf{20,213 (483)} & \textbf{-45.19} \\
\midrule
\multicolumn{6}{l}{\textit{Original embeddings (4096 dimensions)}} \\
$k$NN & 96.41 & 2.12 & 1.47 & 20,669 (530) & -43.95 \\
CRATER (original embeddings) & 96.42 & 2.46 & 1.12 & \textbf{17,190 (573)} & \textbf{-53.39} \\
\bottomrule
\end{tabular}
% ===== END tables/matched_routing.tex =====

\end{table}

With LDA features, $k$NN and CRATER attain \RPLdaKnnAccuracy\% and \RPLdaRiskAccuracy\% correct completion, with cost reductions of \RPLdaKnnSaving\% and \RPLdaRiskSaving\% relative to the reference. CRATER reduces cost by \RPLdaReduction\% relative to $k$NN. Thus both within-representation comparisons give lower cost for CRATER-based routing with similar correct-completion point estimates.

All selected policies meet the validation constraints. On evaluation, both $k$NN policies and the CRATER-based LDA policy exceed the unfinished-response limit; the LDA $k$NN policy also exceeds the correctness-loss limit. Zero Router and the CRATER-based original-embedding policy satisfy the corresponding evaluation constraints at their observed rates.

Cost savings remain stable across capability-based price multipliers when prices reflect differences in active parameter counts. Under a uniform base, both CRATER-based routing versions continue to reduce costs, although the savings are smaller. Across all eight pricing scenarios, the CRATER-based routing versions achieve evaluation correct-completion rates of 94.7\%--96.6\%, compared with 97.3\% for the strong reference (Appendix~\ref{app:routing-price-sensitivity}).
\FloatBarrier
% ===== END routing.tex =====

\subsection{Predicting model behaviors from repeated sampling}
\label{sec:repeated-generation}

Repeated sampling generates multiple responses to the same question. For question $i$ and configuration $m$, each sample is generated under the same settings and a token budget $B$. The quantity pass@$k$ is the probability that at least one of $k$ samples completes correctly within its budget, before any subsequent answer-selection step. We predict this probability across token budgets and sample counts using the fitted CRATER model.

Let $p_{im}(B)$ be the true single-sample correct-completion probability. Under conditionally independent and identically distributed samples, the standard repeated-success formula is

\[
\pi_{im}^{(k)}(B)=1-\{1-p_{im}(B)\}^k.
\]
It follows because all $k$ samples fail to complete correctly with probability $\{1-p_{im}(B)\}^k$. A natural plug-in estimator uses CRATER's single-sample prediction:

\[
\widehat p_{im}(B)=\widehat F_1(B/C\mid x_i,m),
\qquad
\widehat\pi_{im,\mathrm{naive}}^{(k)}(B)
=1-\{1-\widehat p_{im}(B)\}^k.
\]

Directly substituting $\widehat p$ into $1-(1-p)^k$ can yield inaccurate pass@$k$ predictions because of errors in the predicted single-sample probability and residual variation across questions. We therefore use a beta-binomial model to account for both.

Following the beta-binomial approach of \citet{kazdan2025passk} for predicting pass@$k$ from repeated success counts, we model variation in success probabilities across questions through a beta mixing distribution. We use CRATER's predicted single-sample success probability to specify the mean of the beta distribution:

\begin{align}
P_{im}(B)\mid\widehat p_{im}(B)
&\sim\operatorname{Beta}\{\mu_{im}(B)\phi_m,[1-\mu_{im}(B)]\phi_m\},
\label{eq:pass-beta}\\
\logit\mu_{im}(B)
&=a_m+\exp(b_m)\logit\widehat p_{im}(B),\qquad \phi_m>0.
\label{eq:pass-mean}
\end{align}
The mean link corrects systematic differences between predicted and observed single-sample probabilities through monotone logistic recalibration \citep{mishra2022recalibration}; $\phi_m$ controls the residual variation around that mean. The parameters $(a_m,b_m,\phi_m)$ are shared across budgets within each configuration. Repeated samples on a given question share the same latent probability $P_{im}(B)$ and are independent conditional on it.

We estimate the calibration parameters using a beta-binomial count likelihood based on training-question success counts and their fixed CRATER predictions. Because the same samples contribute counts at multiple budgets, the sum of their marginal log likelihoods is a composite likelihood. Integrating the repeated-success probability over the beta distribution gives

\begin{equation}
\widehat\pi_{im}^{(k)}(B)
=1-\frac{\operatorname{Be}\{\widehat\mu_{im}(B)\widehat\phi_m,[1-\widehat\mu_{im}(B)]\widehat\phi_m+k\}}
{\operatorname{Be}\{\widehat\mu_{im}(B)\widehat\phi_m,[1-\widehat\mu_{im}(B)]\widehat\phi_m\}},
\label{eq:bb}
\end{equation}
where $\operatorname{Be}$ is the beta function, and $\widehat\mu_{im}(B)$ and $\widehat\phi_m$ are obtained from the estimated calibration parameters. This prediction averages the repeated-success probability over the conditional beta distribution, incorporating both mean recalibration and residual heterogeneity.

\paragraph{Experimental setup.}
We use the CRATER predictor trained in Section~\ref{sec:results}, keeping its LDA representation, feature scaling, and competing-risks parameters fixed. The beta-binomial parameters are estimated using all ten samples for each of the 1400 training questions and each configuration, together with their fixed single-sample predictions at budgets 128, 256, 512, 1024, 2048, and 4096. The two stages thus use the same training questions. We then evaluate repeated-success predictions on the separate set of 1387 evaluation questions. Appendix~\ref{app:repeated-details} gives the calibration details.

We assess the predictions on the 1387 evaluation questions, using their ten samples as an empirical reference. For question $i$ and configuration $m$, let $D_{im}(B)$ count correct completions within budget $B$ among $R$ evaluation samples. The empirical pass@$k$ estimator is

\begin{equation}
\widehat{\operatorname{pass@}k}_{im}(B)
=1-\frac{\binom{R-D_{im}(B)}{k}}{\binom{R}{k}},\qquad k=1,\ldots,R,
\label{eq:pass}
\end{equation}
where the numerator is zero when $R-D_{im}(B)<k$ \citep{chen2021code}. % <!--ref:chen2021code--><!--anchor:section:2.1-->
Under conditional iid sampling, this estimates the probability of at least one success in $k$ samples. We compare predictions with these estimates for $k=1,\ldots,10$.

Table~\ref{tab:limited} compares the naive plug-in and beta-binomial predictions with the empirical estimates at $k=5$ and $k=10$. Both predictions use the same single-sample predictor, so their comparison assesses mean recalibration and beta mixing jointly. Appendix Table~\ref{tab:passk-complete} reports the complete results for $k=1,\ldots,10$ at all six budgets, including the token-effect ablation below.

\begin{table}[!htbp]
\centering\small
\setlength{\tabcolsep}{4pt}
\caption{Mean pass@5 and pass@10 before (Naive plug-in) and after (Beta-binomial) calibration. Both predictions use CRATER. Calibration uses all ten samples per training question--configuration pair and the fixed Section~\ref{sec:results} predictor. Entries average over 1387 evaluation questions and ten configurations; empirical estimates use ten samples per question--configuration pair. Within each budget--$k$ pair, green bold entries indicate the prediction closer to the empirical estimate in absolute difference; red entries indicate the prediction farther from the empirical estimate.}
\label{tab:limited}
% ===== BEGIN tables/passk_beta_application.tex =====

\begingroup
\newcommand{\closer}[1]{\textcolor[HTML]{4D8067}{\textbf{#1}}}
\newcommand{\farther}[1]{\textcolor[HTML]{C87575}{#1}}

\begin{tabular}{ccccccc}
\toprule
Budget & \multicolumn{3}{c}{pass@5} & \multicolumn{3}{c}{pass@10} \\
\cmidrule(lr){2-4}\cmidrule(lr){5-7}
(tokens) & Empirical & Naive plug-in & Beta-binomial
         & Empirical & Naive plug-in & Beta-binomial \\
\midrule
128  & 0.003 & \farther{0.053} & \closer{0.012}
     & 0.004 & \farther{0.098} & \closer{0.014} \\
256  & 0.091 & \farther{0.263} & \closer{0.088}
     & 0.105 & \farther{0.406} & \closer{0.103} \\
512  & 0.483 & \farther{0.788} & \closer{0.482}
     & 0.514 & \farther{0.901} & \closer{0.521} \\
1024 & 0.812 & \farther{0.970} & \closer{0.816}
     & 0.838 & \farther{0.993} & \closer{0.842} \\
2048 & 0.920 & \farther{0.991} & \closer{0.905}
     & 0.938 & \farther{0.998} & \closer{0.923} \\
4096 & 0.944 & \farther{0.994} & \closer{0.924}
     & 0.958 & \farther{0.999} & \closer{0.939} \\
\bottomrule
\end{tabular}
\endgroup
% ===== END tables/passk_beta_application.tex =====

\end{table}

Calibration brings the predicted mean pass@10 closer to the empirical estimates at every reported budget. At 512 tokens, for example, the naive plug-in prediction is 0.901, compared with 0.521 after calibration and an empirical estimate of 0.514.

\paragraph{Token-effect ablation.}
To examine whether modeling completion time remains useful after beta-binomial calibration, we use the constant-hazard ablation in~\eqref{eq:ablation-no-token}. It removes both the shared $\log s$ trend and the time-varying local adjustment, retaining question effects and configuration- and cause-specific intercepts. The remaining coefficients are reestimated, and a separate beta-binomial layer is trained using the same 1400 questions, ten samples per question--configuration pair and six budgets as for CRATER. Thus both models receive the same calibration procedure.

Table~\ref{tab:pass-time-ablation} compares the calibrated mean pass@10 predictions with the empirical estimates. Across the 60 budget--$k$ combinations in Appendix Table~\ref{tab:passk-complete}, the average absolute difference between predicted and empirical mean pass@$k$ is 0.00887 for CRATER and 0.06433 without token effects. Thus, the shared beta-binomial calibration used here does not fully compensate for the removal of token effects.

\begin{table}[!htbp]
\centering\small
\caption{Token-effect ablation for mean pass@10, comparing CRATER with the model without token effects. Each model has a separately estimated beta-binomial layer trained on the same 1400 questions and ten samples per question--configuration pair. Entries average over 1387 evaluation questions and ten configurations; empirical estimates use ten samples per pair. In each row, green bold entries indicate the prediction closer to the empirical estimate in absolute difference; light red entries indicate the prediction farther from the empirical estimate.}
\label{tab:pass-time-ablation}
% ===== BEGIN tables/passk_time_ablation.tex =====

\begingroup
\newcommand{\closer}[1]{\textcolor[HTML]{4D8067}{\textbf{#1}}}
\newcommand{\farther}[1]{\textcolor[HTML]{C87575}{#1}}

\begin{tabular}{cccc}
\toprule
Budget & Empirical & \multicolumn{2}{c}{Beta-binomial calibrated} \\
\cmidrule(lr){3-4}
(tokens) & & Without token effects & CRATER \\
\midrule
128  & 0.0035 & \farther{0.0803} & \closer{0.0143} \\
256  & 0.1052 & \farther{0.2192} & \closer{0.1026} \\
512  & 0.5144 & \farther{0.4641} & \closer{0.5210} \\
1024 & 0.8381 & \farther{0.7269} & \closer{0.8423} \\
2048 & 0.9376 & \farther{0.8916} & \closer{0.9228} \\
4096 & 0.9578 & \closer{0.9519}  & \farther{0.9393} \\
\bottomrule
\end{tabular}
\endgroup
% ===== END tables/passk_time_ablation.tex =====

\end{table}

At the 4096-token budget, CRATER’s predicted mean pass@10 is 1.9 percentage points below the empirical estimate. The model without token effects is closer to the empirical estimate at this budget, although CRATER performs better across budgets overall. Appendix~\ref{app:calibration-sensitivity} shows that relaxing the calibration model's parameter sharing across budgets brings mean pass@10 closer to the empirical estimate using the same single-sample predictions from CRATER. This sensitivity analysis helps explain the maximum-budget result, although the more flexible calibration can produce predictions that decrease as the budget increases. 
\FloatBarrier
% ===== END downstream.tex =====

% ===== BEGIN discussion.tex =====
\section{Discussion}\label{sec:discuss}

CRATER jointly models completion time and correctness to assess language-model responses under token-budget constraints. The competing-risks formulation distinguishes the probabilities of correct completion, incorrect completion, and remaining unfinished throughout the observed budget range. Expected token consumption follows from the same distribution, connecting the quality of an answer to the resources needed to obtain it. This common structure allows completed and censored responses to contribute to estimation and supplies consistent probability targets for different generation decisions. Its practical value lies in using one fitted model to describe response behavior and support subsequent resource allocation.

The empirical analysis shows that both question information
and token effects contribute to prediction. Their roles are
complementary: question features describe differences between
problems, while token effects describe how completion rates
vary with the token budget. The fitted hazards further reveal
distinct patterns for correct and incorrect completion across
configurations. In particular, a declining correct-completion
hazard indicates that responses still unfinished have a lower
rate of correct completion as the budget increases, even though
cumulative success can continue to rise. Overall accuracy and average response length do not show how correct- and incorrect-completion rates change with the token budget. CRATER describes these changes separately for each completion type and model configuration.

The two important downstream applications illustrate how this distribution can inform different decisions. Routing uses predicted success and expected consumption to select a configuration and budget before generation. The comparisons with nearest-neighbor predictors show lower costs under the primary price assumptions, although savings depend on relative token prices and validation constraints need not hold on evaluation questions. Most selected actions retain the full cap, so these results do not isolate the benefit of budget selection. Repeated-sampling prediction uses the single-attempt probabilities as inputs to a beta-binomial model that adjusts their mean and accounts for residual heterogeneity. The resulting average pass@$k$ predictions agree more closely with empirical estimates than direct plug-in predictions.

The current analysis has several limitations. The empirical evaluation focuses on mathematical reasoning tasks with verifiable answers under fixed generation settings. Extending the framework to open-ended tasks would require a clearly defined criterion for successful completion and a reliable way to assess it. The model also uses only question information, completion lengths and observed outcomes; it does not incorporate token-level log probabilities produced during decoding. These signals may contain additional information about the evolving response and could support updated predictions as generation proceeds. Because they become available during generation, their use would extend the present prediction setting to decisions informed by a partially generated response.

Future work could improve question representations and test the framework on new tasks, model configurations, and independently collected responses. Evaluating the routing strategies under actual serving costs would also help assess their practical value. Incorporating information observed during decoding could also support decisions on whether to continue an attempt or start another. Speculative decoding, in which a smaller model proposes tokens that a target model verifies \citep{leviathan2023speculative}, offers another setting for studying completion probabilities together with the computational costs of drafting and verification. Finally, adding majority voting or verifier-based answer selection would change the target from obtaining at least one correct response to returning a correct selected answer. Modeling that selection step would connect the present framework more directly to the outcome delivered to a user.
% ===== END discussion.tex =====
% ===== END MANUSCRIPT BODY =====

\FloatBarrier
\begingroup
\small

% ===== BEGIN INLINE BIBLIOGRAPHY =====

% ===== END INLINE BIBLIOGRAPHY =====
\endgroup

\clearpage
\appendix

% ===== BEGIN appendix.tex =====
\begin{center}
{\large\bf SUPPLEMENTARY MATERIAL of\\
``Right, Wrong, or Unfinished: A Competing Risks
Framework for Language-Model Reasoning under
Token Budgets''}
\end{center}

% ===== BEGIN appendix_data.tex =====

\section{Data and generation protocol}
\label{app:generation-records}

The analysis contains 1400 MATH training questions and 1387 evaluation questions. Each question has ten response samples from each of ten configurations, giving 278700 records in total. The training set has 40 questions in each of the 35 subject--difficulty cells. The evaluation set has the same cell counts except for a total shortfall of 13 Level 1 questions. Every question--configuration pair has ten distinct seeds, numbered zero through nine. There are 9523 cap-hit records (3.42\%); no observations end before the cap without a recorded completion. The shortest recorded response has 74 tokens. Table~\ref{tab:data} summarizes evaluation outcomes and Table~\ref{tab:generation-checkpoints} gives the full configuration identifiers.

Normalized question text contains no exact duplicates. A character-five-gram cosine screen at 0.90 identifies 32 cross-partition pairs involving 23 evaluation questions. Text similarity does not establish mathematical equivalence, and this screen cannot exclude mathematical overlap or model pretraining contamination.

\begin{table}[!htbp]
\centering\small
\caption{Recorded responses on the 1387 evaluation questions, with 13870 samples per configuration. Correct and incorrect outcomes are recorded completions strictly before the 4096-token cap; cap-hit records are censored. Mean tokens includes capped responses.}
\label{tab:data}
% ===== BEGIN tables/data_summary.tex =====
\begin{tabular}{lrrrr}
\toprule
Configuration & Correct & Incorrect & Cap hits & Mean tokens \\
\midrule
Qwen2.5 14B & 10806 & 3041 & 23 & 532.4 \\
Qwen2.5 7B & 10294 & 3484 & 92 & 578.0 \\
Qwen3 14B & 12115 & 1625 & 130 & 729.9 \\
Qwen3 8B & 11815 & 1778 & 277 & 789.2 \\
Qwen3.5 35B A3B & 12711 & 211 & 948 & 1080.3 \\
Qwen3.5 9B & 12235 & 285 & 1350 & 1228.7 \\
Qwen3.6 35B A3B & 13152 & 259 & 459 & 1062.0 \\
Gemma4 26B A4B & 13128 & 241 & 501 & 866.8 \\
Gemma4 31B & 13496 & 330 & 44 & 594.8 \\
Gemma4 E4B & 11368 & 2125 & 377 & 1095.6 \\
\bottomrule
\end{tabular}
% ===== END tables/data_summary.tex =====

\end{table}

\begin{table}[!htbp]
\centering\footnotesize
\caption{Generation checkpoints and precision settings. Nine configurations use 4-bit AWQ weight quantization; Gemma4-E4B uses a non-quantized checkpoint with FP16 inference.}
\label{tab:generation-checkpoints}
% ===== BEGIN tables/generation_checkpoints.tex =====
\begin{tabular}{@{}p{.22\linewidth}p{.57\linewidth}p{.15\linewidth}@{}}
\toprule
Configuration & Checkpoint identifier & Precision \\
\midrule
Qwen2.5 14B & \path{qwen/Qwen2.5-14B-Instruct-AWQ} & 4-bit AWQ \\
Qwen2.5 7B & \path{qwen/Qwen2.5-7B-Instruct-AWQ} & 4-bit AWQ \\
Qwen3 14B & \path{qwen/Qwen3-14B-AWQ} & 4-bit AWQ \\
Qwen3 8B & \path{qwen/Qwen3-8B-AWQ} & 4-bit AWQ \\
Qwen3.5 35B A3B & \path{cyankiwi/Qwen3.5-35B-A3B-AWQ-4bit} & 4-bit AWQ \\
Qwen3.5 9B & \path{cyankiwi/Qwen3.5-9B-AWQ-4bit} & 4-bit AWQ \\
Qwen3.6 35B A3B & \path{cyankiwi/Qwen3.6-35B-A3B-AWQ-4bit} & 4-bit AWQ \\
Gemma4 26B A4B & \path{cyankiwi/gemma-4-26B-A4B-it-AWQ-4bit} & 4-bit AWQ \\
Gemma4 31B & \path{cyankiwi/gemma-4-31B-it-AWQ-4bit} & 4-bit AWQ \\
Gemma4 E4B & \path{google/gemma-4-E4B-it} & FP16 \\
\bottomrule
\end{tabular}
% ===== END tables/generation_checkpoints.tex =====

\end{table}

The recovered generation workflow uses a fixed system instruction requesting step-by-step reasoning and a user message containing the original problem. Each tokenizer applies its chat template with an assistant-generation prefix. The prompt is reused across seeds, with temperature 0.7, top-$p$ 0.95 and a maximum of 4096 output tokens. We set thinking mode to off, although a small number of outputs still contain explicit thinking segments or related markers.

We assess completed responses using Math-Verify \citep{kydlicek2025mathverify}, which extracts answers from the generated response and the MATH reference solution and checks their mathematical equivalence. % <!--ref:kydlicek2025mathverify--><!--anchor:section:Architecture-->

% ===== END appendix_data.tex =====

% ===== BEGIN appendix_model.tex =====
\section{Model properties and computation}
\subsection{Penalized estimation}
\label{app:math}

The penalty in~\eqref{eq:full-penalty} acts only on slope changes. The parameterization uses a single intercept $a_{km}$ for each configuration and completion type, with no additional global intercept in $\beta_k$.

To see the fused-lasso correspondence explicitly, let $d_{km,0}=b_{km}$ and $d_{km,r}=b_{km}+\sum_{j=1}^r\gamma_{kmj}$, $r=1,\ldots,K$, denote the slopes of successive intervals. Then

\begin{equation}
\lambda\sum_{k,m}\sum_{j=1}^K|\gamma_{kmj}|
=\lambda\sum_{k,m}\sum_{r=1}^K|d_{km,r}-d_{km,r-1}|.
\label{eq:slope-fusion}
\end{equation}
Thus,~\eqref{eq:full-penalty} is a fused-lasso difference penalty on the slopes of the segments, equal to $\lambda$ times their total variation. A zero difference joins two neighboring linear pieces into a single linear segment. The resulting $\alpha_{km}(s)$ is therefore continuous and piecewise linear, with fusion occurring in its slopes.

For fixed $c$, the negative log-likelihood is convex in these coefficients: integrated hazards are integrals of exponentials of affine functions, and event log-hazard contributions are affine. The slope-fusion penalty is also convex. Thus a point satisfying the first-order optimality conditions is a global minimizer of the finite-dimensional working objective.

The fusion penalty leaves the question coefficients, intercepts, and initial slopes unpenalized, so it does not by itself imply coercivity, existence of a finite minimizer for every data set, or uniqueness. 
The reported fits are finite and satisfy the numerical stopping criterion in Appendix~\ref{app:numerics}. Bootstrap summaries assess stability for these data.

\subsection{Numerical integration and optimization}
\label{app:numerics}

We evaluate the likelihood using six-point Gaussian quadrature on each single-token interval. The factor $s^c$ is evaluated directly at the quadrature nodes, allowing the same procedure to be used for the main specification $c=1$ and the other values of $c$ considered in the sensitivity analysis. 
For the main specification $c=1$, the cumulative hazard also has an analytic expression on each spline interval. Writing the hazard on $[l,r]$ as $u\exp(A+Du)$ gives, for $D\ne0$,

\[
\int_l^r u\exp(A+Du)\,du
=\exp(A)\left[\frac{\exp(Du)(Du-1)}{D^2}\right]_l^r.
\]
When $D=0$, the integral reduces to $\exp(A)(r^2-l^2)/2$.

Prediction integrates $S(u)h_k(u)$ directly on each single-token interval, evaluating the cumulative hazard at the quadrature nodes. Allocating an interval's total event mass in proportion to its integrated cause-specific hazards would be exact only if their relative shares were constant within that interval. Direct integration avoids that approximation. Comparing six- and ten-point quadrature checks probability accuracy; conservation and monotonicity are checked separately, without renormalizing the predictions.

Let $g(\theta)$ be the differentiable negative log likelihood and let $w$ be zero except for weight $\lambda$ on each hinge coefficient. The stopping residual is
$\left\|\theta-\operatorname{soft}\{\theta-\nabla g(\theta),w\}\right\|_2$ where $\operatorname{soft}(z,w)=\operatorname{sign}(z)(|z|-w)_+$.
A proximal Newton method solves the local quadratic problem with the exact absolute-value penalty; split-variable optimization provides a fallback. Training requires a stopping residual of at most $5\times10^{-8}$.

CRATER and all component ablations use this likelihood, derivative and prediction implementation. For the two ablations without token effects, the analytic probabilities are $S(s)=\exp\{-(A_1+A_2)s\}$ and $F_k(s)=A_k\{1-S(s)\}/(A_1+A_2)$, where $A_k=\exp(x_i^\top\beta_k+a_{km})$ when question effects are retained and $A_k=\exp(a_{km})$ otherwise. These provide an additional check of the numerical predictions.

After training CRATER, we check the objective and gradient against analytic integration; both discrepancies are below $10^{-15}$.

\FloatBarrier

\subsection{Orthogonal reparameterization for original embeddings}
\label{app:embedding-reparameterization}

For the extension in Remark~\ref{rem:original-embeddings}, let $\mathbf X\in\mathbb R^{n\times d}$ contain one standardized embedding row per training question. Training means and standard deviations are also used for new questions. The estimation criterion is

\[
\mathcal L(\theta)+\lambda\sum_{k,m,j}|\gamma_{kmj}|
+\frac{\rho}{2}\sum_{k=1}^2\|\beta_k\|_2^2.
\]
The ridge penalty acts on coefficients in these standardized original coordinates. Intercepts and initial time-adjustment slopes remain unpenalized.

\citet{hastie2004quadratic} show that orthogonal reparameterization permits exact estimation of quadratically penalized linear-predictor models in a smaller coordinate space, including logistic and Cox regression. Their construction uses the singular value decomposition. We apply the same principle to the question coefficients, constructing the orthogonal coordinates by QR decomposition and a subsequent rotation; the time-effect parameterization and its penalty are unchanged. % <!--ref:hastie2004quadratic--><!--anchor:section:3.1, Theorem 1-->

When $d>n$, an economic QR decomposition gives $\mathbf X^\top=QR$, where $Q\in\mathbb R^{d\times n}$ has orthonormal columns and its column space contains the row space of $\mathbf X$. Write $\beta_k=Q\eta_k+v_k$ with $Q^\top v_k=0$. Then
\[
\mathbf X\beta_k=\mathbf XQ\eta_k,
\qquad
\|\beta_k\|_2^2=\|\eta_k\|_2^2+\|v_k\|_2^2.
\]
All training likelihood contributions depend on the question coefficients through $\mathbf X\beta_k$, including repeated response samples and configurations. Thus $v_k$ cannot improve the likelihood, and $\rho>0$ sets it to zero at any minimizer. Optimization over $\eta_k$ is therefore equivalent to optimization over the original ridge-penalized coefficients. When $d\leq n$, we take $Q=I_d$.

To obtain orthogonal design columns, set $A=\mathbf XQ$ and diagonalize $A^\top A/n=UDU^\top$, with $U$ orthogonal and $D$ diagonal. Define $Q_*=QU$ and $H=\mathbf XQ_*$. In the coordinates $\beta_k=Q_*\xi_k$,

\[
H^\top H/n=D,\qquad
\mathbf X\beta_k=H\xi_k,\qquad
\|\beta_k\|_2^2=\|\xi_k\|_2^2.
\]
All eigenvector directions from this diagonalization are retained, without eigenvalue truncation or whitening. Consequently, the transformation preserves both the likelihood and the ridge penalty on the retained coefficient space. Ridge provides statistical regularization; the orthogonal coordinates facilitate computation.

For a new standardized embedding $x$, the question effect is $(Q_*^\top x)^\top\widehat\xi_k=x^\top\widehat\beta_k$. Hence, this implementation also preserves predictions from the corresponding model in original coordinates. The scaler and transformation are learned separately from the training questions in each split and then fixed for prediction.

% ===== END appendix_model.tex =====

% ===== BEGIN appendix_fits.tex =====
\section{Estimation settings and supplementary fits}
\label{app:estimation-supplement}

\subsection{Exploratory comparison of question representations}
\label{app:representation-pilot}

To choose how to reduce the embeddings, we first examine whether the resulting coordinates distinguish known aspects of question structure. An initial pilot compares unreduced embeddings, principal component analysis (PCA) with 32 or 64 components, and linear discriminant analysis (LDA), using Qwen and BERT encoders. We summarize separation by subject and difficulty with mean silhouette scores based on cosine distance; larger scores indicate more compact and better-separated label groups \citep{rousseeuw1987}. % <!--ref:rousseeuw1987--><!--anchor:section:Abstract-->
For Qwen embeddings, LDA gives scores of 0.9442 for subject and 0.8550 for difficulty, compared with scores near zero for the unreduced and PCA representations (Table~\ref{tab:silhouette}). This pilot motivates using LDA to form a compact representation guided by subject and difficulty labels.

\begin{center}
\begin{minipage}{\textwidth}
\centering\small
\captionof{table}{Mean silhouette scores for separation by subject and difficulty. Dimension-reduction training and scoring use the same 5,000 MATH benchmark test questions. Subject and difficulty LDA projections are learned separately, with six and four coordinates, respectively.}
\label{tab:silhouette}
% ===== BEGIN tables/pilot_silhouette.tex =====
\begin{tabular}{llrr}
\toprule
Encoder & Reduction & Subject & Difficulty \\
\midrule
Qwen3-Embedding-8B & None & 0.0205 & -0.0049 \\
 & PCA-64 & 0.0294 & -0.0096 \\
 & PCA-32 & 0.0319 & -0.0137 \\
 & LDA & \textbf{0.9442} & \textbf{0.8550} \\
\midrule
BERT-base & None & -0.0653 & -0.0366 \\
 & PCA-64 & -0.0484 & -0.0250 \\
 & PCA-32 & -0.0554 & -0.0282 \\
 & LDA & \textbf{0.3625} & \textbf{0.0170} \\
\bottomrule
\end{tabular}
% ===== END tables/pilot_silhouette.tex =====

\end{minipage}
\end{center}

LDA emphasizes between-class separation relative to within-class variation \citep{fisher1936}. % <!--ref:fisher1936--><!--anchor:page:179-181-->
Because the pilot uses the same labels to learn its LDA projections and assess their separation, its scores describe the fitted label structure. They provide a motivation for the working representation; component contributions are assessed in Section~\ref{sec:ablation}, with evaluation-sample probability diagnostics in Appendix~\ref{app:state-probabilities}.

\subsection{Representation, knots, and validation}
\label{app:development-design}
The main CRATER specification uses the ten-dimensional question representation obtained by dimension reduction in Section~\ref{sec:model-specification}. Both LDA maps are learned from the 1400 training questions and held fixed across the three internal validation splits. Thus the labels of internal-validation questions contribute to representation learning, while evaluation-question labels do not. Each split contains 1120 questions for outcome-model training and 280 for validation, stratified by subject and difficulty. Feature centering and scaling are estimated from the 1120 training questions. All configurations and response samples of a question remain together.

The time adjustment allows slope changes at eighteen prespecified knot locations, expressed in output tokens as follows:
128,\ 192,\ 256,\ 384,\ 512,\ 768,\ 1024,\ 1280,\ 1536,
1792,\ 2048,\ 2304,\ 2560,\ 2816,\ 3072,\ 3328,\ 3584,\ 3840.
Dividing by 4096 gives the normalized $\kappa_j$. This grid was fixed before the reported comparisons and places more knots early in generation.

For each fixed $c$, we select $\lambda$ from 19 values between $10^{-7}$ and $10^{-3}$ by minimizing mean validation IBS across three splits. With $c=1$, the selected value is $\lambda=10^{-4}$. We then train the final model on all 1400 training questions.

\subsection{Shared time trend and sensitivity analysis}
\label{app:power-sensitivity}

We report sensitivity results for $c\in\{0,1,2,4\}$ in~\eqref{eq:hazard} to assess how strongly predictive performance depends on the shared time trend. The coefficient $c$ determines the power-law component $s^c$, while $\alpha_{km}(s)$ allows configuration- and cause-specific departures. The case $c=0$ omits the shared log-time term.

The LDA representation, question partition, knot locations, and evaluation procedure remain fixed across comparisons. All values of $c$ use the same numerical integration procedure described in Appendix~\ref{app:numerics}.

\begin{center}
\begin{minipage}{\textwidth}
\centering\small
\captionof{table}{Sensitivity to the shared time coefficient $c$. For each reported $c$, minimizing mean validation IBS across three training-sample splits over 19 candidate penalties from $10^{-7}$ to $10^{-3}$ selects $\lambda=10^{-4}$. Evaluation IBS is computed after training on all training questions with the selected penalty.}
\label{tab:c}
% ===== BEGIN tables/section4_c_sensitivity.tex =====
\begin{tabular}{rrr}
\toprule
$c$ & Validation IBS & Evaluation IBS \\
\midrule
0 & 0.073383 & 0.091111 \\
1 & 0.073382 & 0.091030 \\
2 & 0.073383 & 0.091021 \\
4 & 0.073390 & 0.091168 \\
\bottomrule
\end{tabular}
% ===== END tables/section4_c_sensitivity.tex =====

\end{minipage}
\end{center}

Table~\ref{tab:c} shows that evaluation IBS ranges from 0.09102 to 0.09117 across the reported values. In particular, $c=1$ and $c=2$ give nearly identical validation and evaluation scores. The comparable performance at $c=0$ also indicates that the shared log-time term is not essential for predictive accuracy within this flexible model family.

The local adjustment can compensate for part of the shared trend, allowing different values of $c$ to yield similar prediction scores. We retain $c=1$ as a fixed reference for subsequent estimation and ablation analyses, with $\alpha_{km}(s)$ describing departures from the shared $\log s$ trend. The sensitivity results support robustness of aggregate predictive performance to this choice; they do not identify a unique time exponent.

\subsection{Bootstrap summaries}
Parameter summaries use 200 bootstrap samples of training questions drawn within subject--difficulty strata, retaining all configurations and response samples of every selected question. The LDA map, feature scaling, knots, $c$, and selected penalty remain fixed. We report bootstrap standard deviations for the coefficients and use the 2.5th and 97.5th percentiles of the bootstrap curves to construct pointwise 95\% bands. These summaries describe variability conditional on the representation and selected specification, rather than simultaneous coverage or uncertainty from repeating model selection.

Score and routing comparisons use 4000 paired bootstrap samples of questions within subject--difficulty strata, preserving each stratum's size. All compared methods share the same sampled questions; predictors and policies remain fixed. Training and evaluation scores are resampled separately. These summaries quantify variation across questions within the observed strata, without retraining or correcting training-score optimism.

\subsection{Time adjustments}
\begin{figure}[!ht]
\centering\includegraphics[width=.80\textwidth,height=.70\textheight,keepaspectratio]{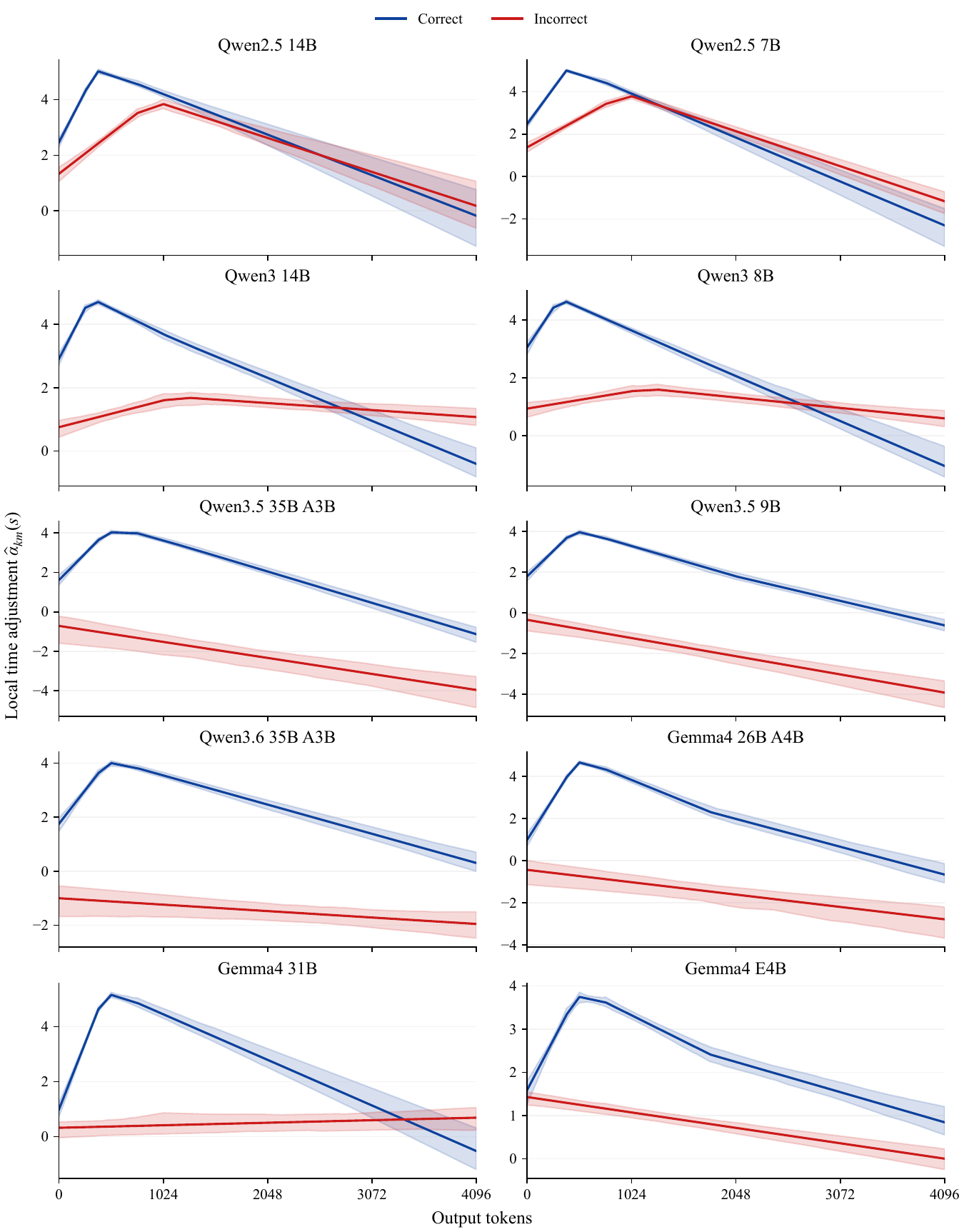}
\caption{Fitted time adjustments for correct and incorrect completion across configurations. Shaded bands are pointwise 95\% percentile intervals from 200 stratified question bootstrap samples under the fixed representation and specification. The adjustments describe departures from the shared $\log s$ trend.}
\label{fig:alpha}
\end{figure}

Figure~\ref{fig:alpha} shows the time adjustments underlying the hazards in Figure~\ref{fig:hazards}. They describe departures from the shared $\log s$ trend. At centered question features, $h_k(s\mid 0,m)=s\exp\{\alpha_{km}(s)\}$, so a peak in the adjustment need not coincide with a peak in the hazard.

\FloatBarrier

\subsection{Training-sample state probabilities by difficulty}
\label{app:training-states}

\begin{figure}[!htbp]
\centering
\includegraphics[width=\textwidth]{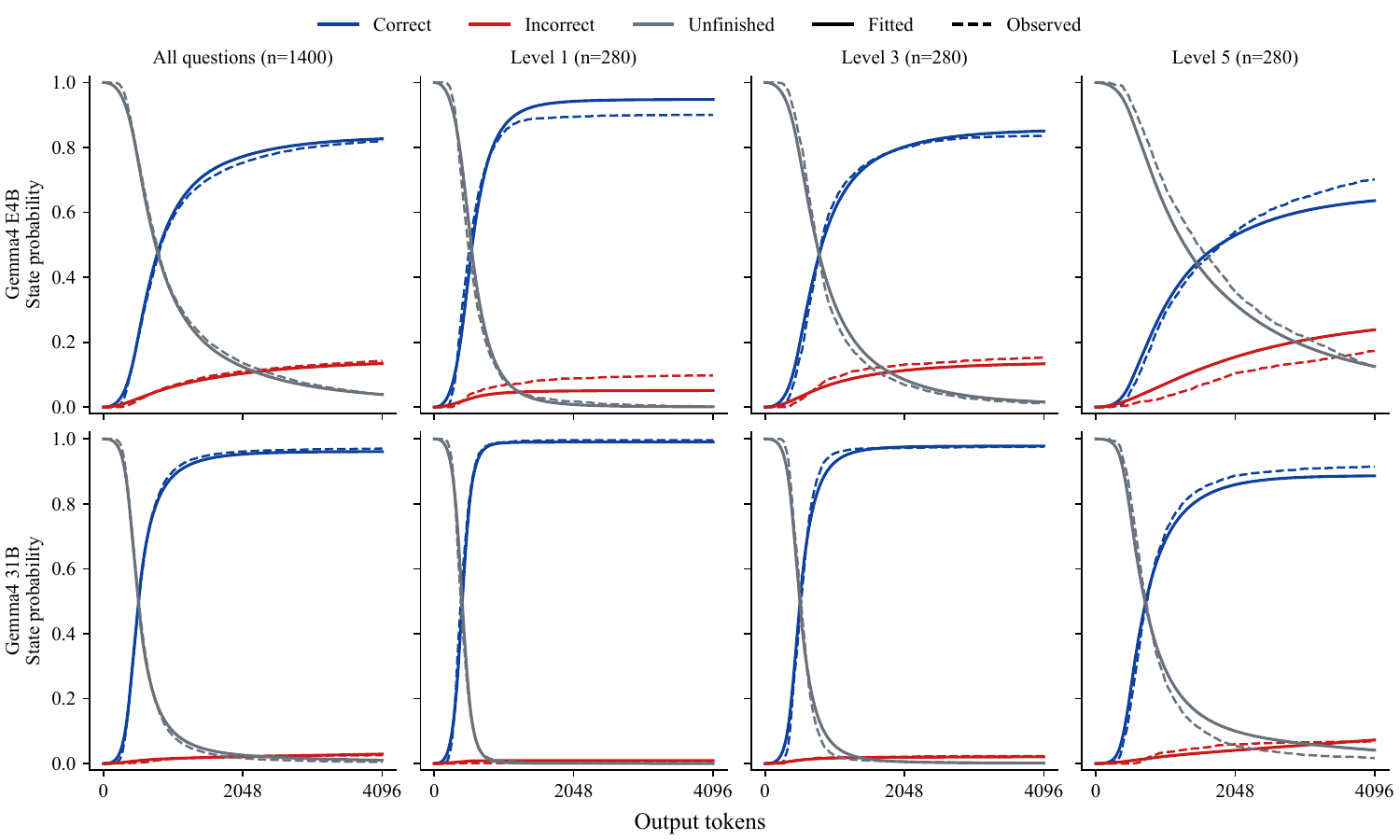}
\caption{Fitted and observed state probabilities on training questions for Gemma4-E4B (top) and Gemma4-31B (bottom), overall and at Levels 1, 3 and 5. Solid curves average fitted probabilities within each group; dashed curves show observed proportions across all associated response samples. Panel headings give question counts.}
\label{fig:states-training-stratified}
\end{figure}

Figure~\ref{fig:states-training-stratified} supplements the pooled training-sample curves in Figure~\ref{fig:states-training} with Levels 1, 3 and 5, using the same fitted model. Agreement is less uniform within difficulty groups: for Gemma4-E4B, fitted correct-completion probabilities at longer budgets exceed the observed proportions at Level 1 but fall below them at Level 5, with corresponding discrepancies in incorrect completion. Gemma4-31B shows closer agreement at Levels 1 and 3, while discrepancies remain at Level 5. Thus similar pooled curves can conceal differences across difficulty groups, even within the training sample.

\FloatBarrier

\subsection{Evaluation-sample state probabilities}
\label{app:state-probabilities}

Within each configuration and question group, solid curves show average predicted probabilities, while dashed curves show the observed proportions of correct completion, incorrect completion and unfinished responses at each token budget.

\begin{figure}[!htb]
\centering
\includegraphics[width=\textwidth]{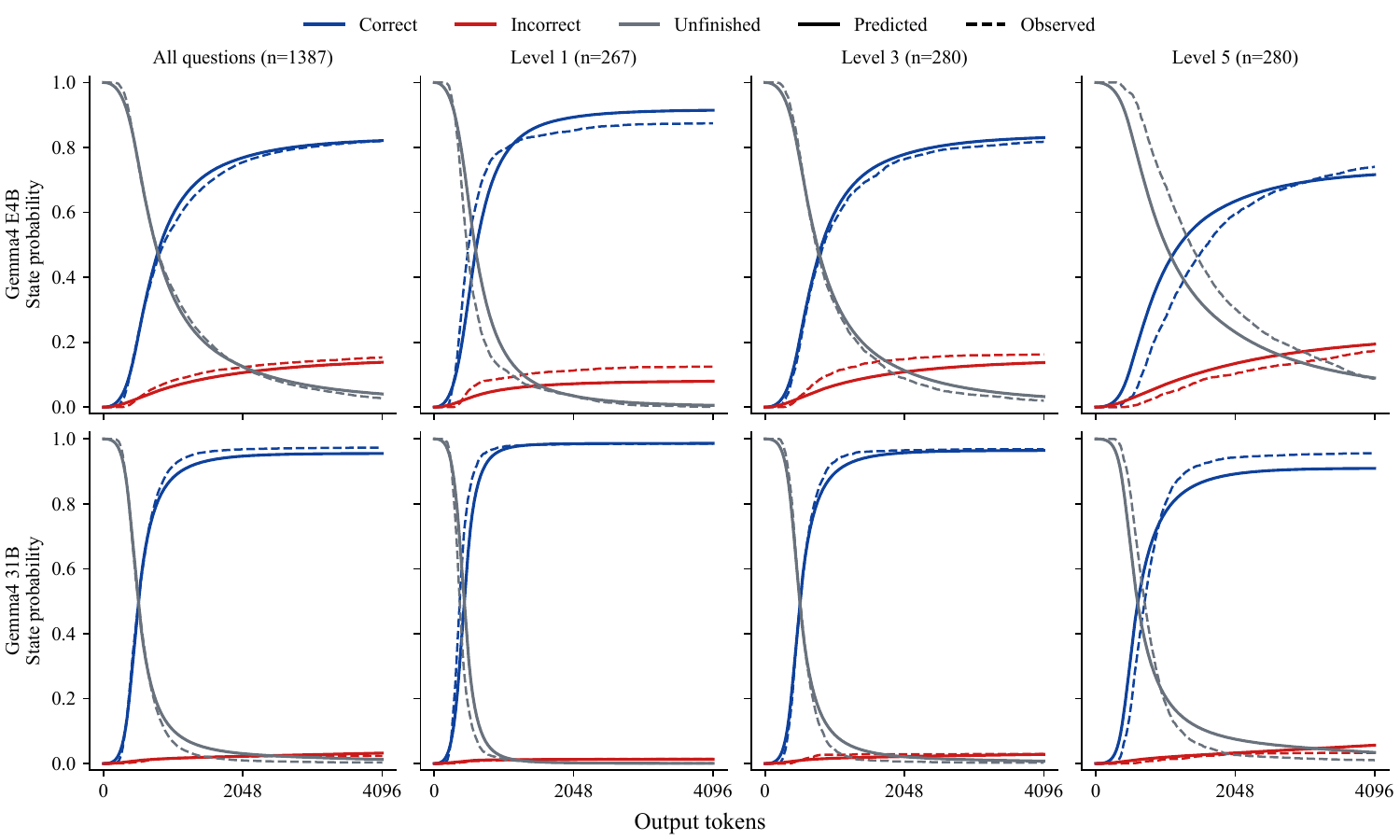}
\caption{Predicted and observed state probabilities for Gemma4-E4B (top) and Gemma4-31B (bottom), across all evaluation questions and questions at Levels 1, 3 and 5 (columns, left to right). Solid curves average predictions from the same fitted model within each group; dashed curves show observed proportions across all response samples in that group. Panel headings give question counts. Colors distinguish correct completion, incorrect completion and no recorded completion. Overall agreement can coexist with differences in completion timing within difficulty groups.}
\label{fig:states}
\end{figure}

\begin{figure}[!htb]
\centering\includegraphics[width=.91\textwidth]{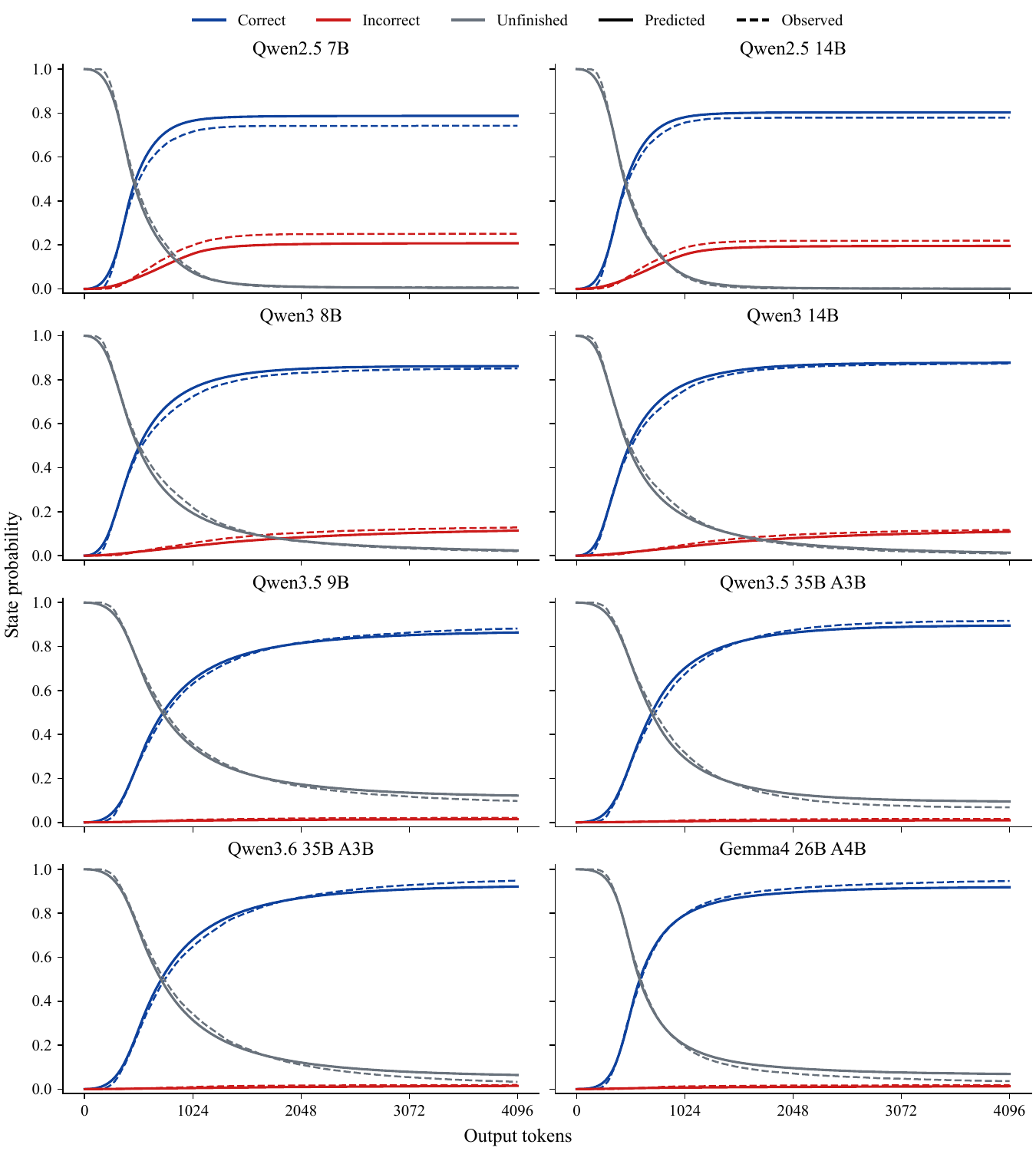}
\caption{Predicted and observed state probabilities for the eight configurations omitted from Figure~\ref{fig:states}. Solid curves average predictions over evaluation questions; dashed curves show observed proportions across all response samples on those questions. Colors distinguish correct completion, incorrect completion and no recorded completion.}
\label{fig:states-additional}
\end{figure}

Figure~\ref{fig:states} illustrates a relatively weaker configuration, Gemma4-E4B, and a stronger configuration, Gemma4-31B, whose observed correct-completion rates at the maximum budget are 82.0\% and 97.3\%, respectively. The rows correspond to these two configurations, and the columns show all evaluation questions and questions at Levels 1, 3 and 5. These levels assess whether overall agreement persists across low, intermediate and high difficulty. All panels use the same fitted model. Figure~\ref{fig:states-additional} provides pooled evaluation curves for the remaining eight configurations.

The model captures the earlier accumulation and higher final proportion of correct completions for Gemma4-31B relative to Gemma4-E4B. Agreement is less uniform across difficulty levels: in both configurations, correct completions initially accumulate earlier than predicted for Level 1 questions and later than predicted for Level 5 questions. Across all ten configurations at the maximum budget, average predicted correct completion is close to the observed proportion, while incorrect completion is underestimated and the unfinished proportion is overestimated. Thus agreement in overall correct completion can coexist with discrepancies in completion timing and the other outcome probabilities.

% ===== END appendix_fits.tex =====

% ===== BEGIN appendix_routing.tex =====
\section{Routing methods and sensitivity analyses}
\label{app:routing-supplement}
% ===== BEGIN routing_appendix.tex =====
\subsection{Routing predictors and policy selection}
\label{app:pricedrouting}

\paragraph{Prediction inputs.}
We consider the CRATER-based router with and without dimension reduction. With dimension reduction, it uses the ten-dimensional question representation and competing-risks specification from Section~\ref{sec:results}, with $c=1$ and the slope-fusion penalty $\lambda=10^{-4}$. Without dimension reduction, it retains all 4096 embedding coordinates and adds ridge regularization to the question coefficients, while keeping the same time-effect specification. The resulting penalty is
\[
\lambda\sum_{k,m,j}|\gamma_{kmj}|+\frac{\rho}{2}\sum_k\|\beta_k\|_2^2,
\]
added to the mean negative log-likelihood. Intercepts and initial slopes remain unpenalized. Mean validation IBS selects $(\lambda,\rho)$ from $\{10^{-5},3\times10^{-5},10^{-4},3\times10^{-4},10^{-3}\}\times\{10^{-6},10^{-5},\ldots,10^2\}$, giving $(3\times10^{-4},1)$.

Under each representation, the CRATER-based router is compared with a $k$NN predictor using the same question features. The dimension-reduction map is learned from the 1400 training questions and held fixed across validation splits, as described in Appendix~\ref{app:development-design}. For the CRATER specifications, feature centering and scaling use the 1120 training questions within each split; final predictors are trained on all 1400 training questions. Both CRATER specifications compute expected token consumption by trapezoidal integration of survival at every integer token budget.

$k$NN without dimension reduction uses cosine distance; $k$NN with dimension reduction uses Euclidean distance after coordinate standardization. Each neighbor contributes its mean correct-by-$B$ indicator and $\min(\widetilde T,B)$ over ten response samples for every configuration and candidate cap $B$. Probability and consumption neighbor counts are selected separately from $\{5,10,20,50,100,200\}$ using mean validation Bernoulli Brier score and squared error of consumption divided by its cap, averaged over the 80 actions. For a fixed question, configuration and budget, let $p$ be the predicted success probability and $y_r\in\{0,1\}$ the observed success indicator for response sample $r$, with $\bar y=R^{-1}\sum_{r=1}^R y_r$ and $R=10$. Since $y_r^2=y_r$, the average Brier score across response samples is
\[
\frac{1}{R}\sum_{r=1}^R(p-y_r)^2
=p^2-2p\bar y+\bar y.
\]
 Both representations select $k=20$ for both targets. Final predictions use neighbors among all 1400 training questions. Without dimension reduction, the two methods share the original embeddings, with cosine normalization for $k$NN and coordinate standardization for CRATER.

Zero Router uses empirical outcome proportions and mean consumption over the pooled validation records for each of the 80 fixed actions. It requires no question representation.

\paragraph{Policy selection.}
Policy selection pools validation predictions from the three splits, weighting each question--split record equally. For numerical scaling, let $C_0$ be the mean observed cost of full-cap Gemma4-31B on the 1400 training questions. Each question-aware strategy constructs $\widehat g_\eta$ at $\eta=0$ and 81 logarithmically spaced values between $10^{-4}/C_0$ and $10^2/C_0$. Together with the strong-reference fallback, this gives 83 candidate rules, indexed by $j$ and denoted $g_j$. For mixture probabilities $v_j$, the action probabilities in Section~\ref{sec:routing} are $\pi_a(x)=\sum_jv_j\ind\{g_j(x)=a\}$. Let $\bar a_j$, $\bar w_j$, $\bar u_j$, and $\bar c_j$ be the validation averages of correct completion, incorrect completion, unfinished responses and cost for candidate rule $j$. With reference averages $a_0$, $w_0$, and $u_0$, mixture weights solve

\begin{align}
\min_{v_j\geq0}\quad &\sum_jv_j\bar c_j,\notag\\
\text{subject to}\quad
&\sum_jv_j\bar a_j\geq a_0-\delta,\qquad
\sum_jv_j\bar w_j\leq w_0+\delta/2,\notag\\
&\sum_jv_j\bar u_j\leq u_0+\delta,\qquad
\sum_jv_j=1.
\label{eq:routing-mixture}
\end{align}
We set $\delta=0.01$. This program minimizes validation cost over mixtures of the constructed candidate rules. Zero Router solves the same linear program over the 80 fixed actions. A question-aware mixture samples a rule with probabilities $v_j$ and applies it to the question; Zero Router samples an action from the same fixed distribution for every question. Evaluation averages over these probabilities, with one action assigned to every question.

With $\delta=0.01$, Zero Router mixes full-cap Gemma4-31B, Qwen3.6-A3B, and Qwen2.5-14B with probabilities 0.7553, 0.2286, and 0.0162. Restricting it to the ten full-cap actions gives the same result. All five selected policies satisfy the validation constraints. On evaluation, Zero Router and the CRATER-based policy without dimension reduction satisfy the corresponding reference-relative constraints. Both $k$NN policies and the CRATER-based policy with dimension reduction exceed the unfinished-response limit; the $k$NN policy with dimension reduction also exceeds the correctness-loss limit. No evaluation outcomes are used to alter these policies.

\FloatBarrier
% ===== END routing_appendix.tex =====

% ===== BEGIN routing_price_sensitivity.tex =====
\subsection{Sensitivity to pricing assumptions}
\label{app:routing-price-sensitivity}

\begin{figure}[!htbp]
\centering
\includegraphics[width=\textwidth]{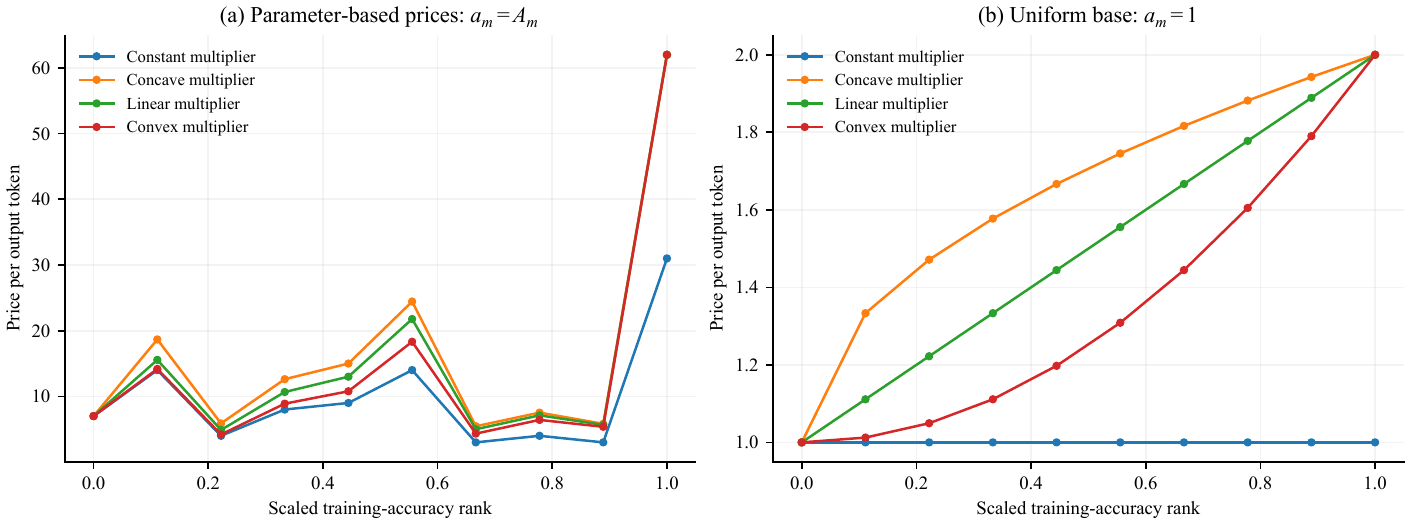}
\caption{Price scenarios ordered by training correct-completion rank. Prices combine either the actual active parameter count (left) or a common base of one (right) with four capability-based price multipliers. Removing parameter-count differences changes the relative prices more substantially than changing the shape of the price multiplier.}
\label{fig:routing-price-sensitivity}
\end{figure}

We examine whether the routing results depend on differences in active parameter counts or on the assumed capability-based price multiplier. Let $q_m=(r_m-1)/(M-1)$, where $r_m$ is configuration $m$'s ascending rank in training correct-completion rate, with average ranks for ties. We consider eight price scenarios,
\[
p_m=a_m g(q_m),\qquad
a_m\in\{A_m,1\},\qquad
g(q)\in\{1,\ 1+\sqrt q,\ 1+q,\ 1+q^2\}.
\]
The first base retains the active parameter counts in Table~\ref{tab:routing-prices}; the second assigns a common base price to all configurations. The four choices of $g(q)$ are constant, concave, linear, and convex functions of capability rank. The main analysis uses $a_m=A_m$ and $g(q)=1+q$. These shapes describe how the price multiplier changes with capability rank; the resulting parameter-based prices need not be monotone in capability (Figure~\ref{fig:routing-price-sensitivity}).

All four question-dependent predictors, their preprocessing, and their training parameters remain fixed at the versions used in Table~\ref{tab:routing}. For each price scenario, we reselect the five adaptive policies, including Zero Router, on the same validation records using the same 80 actions and quality constraints as in Appendix~\ref{app:pricedrouting}. The candidate multiplier grid is normalized by the strong reference's training mean cost under that scenario. Policies are then held fixed for evaluation. All 40 selected policies satisfy the validation constraints. The main-price scenario reproduces Table~\ref{tab:routing}.

When base prices are proportional to active parameter counts, the ranking of routing methods by mean cost changes little across the choices of capability-based price multiplier (Table~\ref{tab:routing-price-actual}). Zero Router saves 19.97--20.61\% relative to the strong reference, compared with 42.86--44.92\% for original-embedding $k$NN and 53.39--54.12\% for CRATER using original embeddings. The latter reduces cost by a further 16.70--19.07\% relative to its matched $k$NN comparator and satisfies all three evaluation quality bounds in each scenario. CRATER with LDA features also reduces cost relative to LDA $k$NN, but exceeds the unfinished-response bound in the three scenarios where the price multiplier increases with capability rank.

With a uniform base, cost savings shrink substantially (Table~\ref{tab:routing-price-uniform}). CRATER using original embeddings saves 5.55--11.73\% relative to the strong reference, and its cost advantage over the matched $k$NN predictor becomes small or reverses. With the convex price multiplier it costs 1.46\% more than $k$NN. Both CRATER-based routing versions exceed the allowed evaluation correctness loss in all four uniform-base scenarios, despite satisfying the validation constraints. These cost reductions therefore do not demonstrate savings while maintaining the specified evaluation quality. The results support stability across the price-multiplier functions under parameter-based pricing, while showing that the magnitude of savings depends strongly on differences in base prices.

\begin{table}
\centering\small
\setlength{\tabcolsep}{3pt}
\caption{Routing sensitivity with the actual parameter base, $p_m=A_m g(q_m)$. Results average 1387 evaluation questions and ten response samples per question, integrating over policy randomization. The strong reference has 97.304\% correct, 2.379\% incorrect and 0.317\% unfinished responses; its mean cost appears within each block. Negative values in the final column denote cost reductions relative to that reference under the same prices. Parentheses report SDs from 4000 paired question bootstrap samples within subject--difficulty strata, with predictors and policies fixed. CRATER without dimension reduction uses ridge regularization.}
\label{tab:routing-price-actual}
% ===== BEGIN tables/routing_price_actual.tex =====
\begin{tabular}{llrrrrr}
\toprule
& Dimension & \multicolumn{3}{c}{Outcome (\%)} & Mean cost & Cost saving \\
Method & reduction & Correct & Incorrect & Unfinished & (SD) & vs. strong (\%) \\
\midrule
\multicolumn{7}{l}{\textit{Constant multiplier: $g(q_m)=1$}} \\
\multicolumn{7}{l}{\quad Strong-reference mean cost: 18,439} \\
Zero Router & --- & 96.396 & 2.601 & 1.003 & 14,757 (178) & -19.97 \\
$k$NN & With & 96.247 & 2.213 & 1.541 & 13,426 (222) & -27.19 \\
CRATER & With & 96.551 & 2.150 & 1.299 & 9,812 (247) & -46.79 \\
$k$NN & Without & 96.538 & 2.076 & 1.385 & 10,537 (264) & -42.86 \\
CRATER & Without & 96.381 & 2.455 & 1.164 & 8,527 (283) & -53.75 \\
\midrule
\multicolumn{7}{l}{\textit{Concave multiplier: $g(q_m)=1+\sqrt{q_m}$}} \\
\multicolumn{7}{l}{\quad Strong-reference mean cost: 36,878} \\
Zero Router & --- & 96.396 & 2.601 & 1.003 & 29,412 (355) & -20.24 \\
$k$NN & With & 96.241 & 2.263 & 1.496 & 26,841 (445) & -27.22 \\
CRATER & With & 96.446 & 2.105 & 1.449 & 19,066 (493) & -48.30 \\
$k$NN & Without & 96.406 & 2.125 & 1.469 & 20,845 (529) & -43.48 \\
CRATER & Without & 96.388 & 2.434 & 1.177 & 16,975 (568) & -53.97 \\
\midrule
\multicolumn{7}{l}{\textit{Linear multiplier: $g(q_m)=1+q_m$; main setting}} \\
\multicolumn{7}{l}{\quad Strong-reference mean cost: 36,878} \\
Zero Router & --- & 96.423 & 2.578 & 0.999 & 29,362 (354) & -20.38 \\
$k$NN & With & 96.269 & 2.253 & 1.477 & 26,928 (444) & -26.98 \\
CRATER & With & 96.402 & 2.110 & 1.488 & 20,213 (483) & -45.19 \\
$k$NN & Without & 96.406 & 2.125 & 1.469 & 20,669 (530) & -43.95 \\
CRATER & Without & 96.419 & 2.463 & 1.118 & 17,190 (573) & -53.39 \\
\midrule
\multicolumn{7}{l}{\textit{Convex multiplier: $g(q_m)=1+q_m^2$}} \\
\multicolumn{7}{l}{\quad Strong-reference mean cost: 36,878} \\
Zero Router & --- & 96.423 & 2.578 & 0.999 & 29,278 (353) & -20.61 \\
$k$NN & With & 96.285 & 2.256 & 1.459 & 26,887 (445) & -27.09 \\
CRATER & With & 96.385 & 2.121 & 1.494 & 19,798 (488) & -46.31 \\
$k$NN & Without & 96.333 & 2.125 & 1.543 & 20,311 (531) & -44.92 \\
CRATER & Without & 96.364 & 2.463 & 1.173 & 16,919 (576) & -54.12 \\
\bottomrule
\end{tabular}
% ===== END tables/routing_price_actual.tex =====

\end{table}

\begin{table}[p]
\centering\small
\setlength{\tabcolsep}{3pt}
\caption{Routing sensitivity with a uniform base, $p_m=g(q_m)$. Evaluation averaging and method labels follow Table~\ref{tab:routing-price-actual}; the same strong-reference outcome rates apply, with scenario-specific mean costs shown within each block. Parentheses report SDs from the same 4000 paired, subject--difficulty-stratified question bootstrap samples, holding predictors and policies fixed. Negative values in the final column denote cost reductions relative to the strong reference under the same prices. In the constant-multiplier block all configurations have unit price, so cost equals the number of output tokens consumed.}
\label{tab:routing-price-uniform}
% ===== BEGIN tables/routing_price_uniform.tex =====
\begin{tabular}{llrrrrr}
\toprule
& Dimension & \multicolumn{3}{c}{Outcome (\%)} & Mean cost & Cost saving \\
Method & reduction & Correct & Incorrect & Unfinished & (SD) & vs. strong (\%) \\
\midrule
\multicolumn{7}{l}{\textit{Constant multiplier: $g(q_m)=1$}} \\
\multicolumn{7}{l}{\quad Strong-reference mean cost: 595} \\
Zero Router & --- & 96.607 & 2.564 & 0.829 & 585 (6) & -1.66 \\
$k$NN & With & 95.836 & 2.974 & 1.189 & 587 (7) & -1.29 \\
CRATER & With & 96.022 & 2.723 & 1.254 & 570 (6) & -4.23 \\
$k$NN & Without & 96.217 & 2.885 & 0.898 & 577 (6) & -3.02 \\
CRATER & Without & 96.000 & 2.893 & 1.106 & 562 (7) & -5.55 \\
\midrule
\multicolumn{7}{l}{\textit{Concave multiplier: $g(q_m)=1+\sqrt{q_m}$}} \\
\multicolumn{7}{l}{\quad Strong-reference mean cost: 1,190} \\
Zero Router & --- & 96.268 & 3.146 & 0.586 & 1,157 (12) & -2.77 \\
$k$NN & With & 95.909 & 3.333 & 0.759 & 1,140 (14) & -4.19 \\
CRATER & With & 94.688 & 4.632 & 0.680 & 1,080 (14) & -9.21 \\
$k$NN & Without & 95.934 & 3.262 & 0.804 & 1,104 (14) & -7.24 \\
CRATER & Without & 95.408 & 3.580 & 1.012 & 1,072 (16) & -9.90 \\
\midrule
\multicolumn{7}{l}{\textit{Linear multiplier: $g(q_m)=1+q_m$}} \\
\multicolumn{7}{l}{\quad Strong-reference mean cost: 1,190} \\
Zero Router & --- & 96.353 & 3.092 & 0.555 & 1,156 (12) & -2.84 \\
$k$NN & With & 95.770 & 3.500 & 0.730 & 1,128 (14) & -5.16 \\
CRATER & With & 95.294 & 4.051 & 0.655 & 1,072 (15) & -9.91 \\
$k$NN & Without & 95.689 & 3.546 & 0.765 & 1,076 (14) & -9.57 \\
CRATER & Without & 95.618 & 3.515 & 0.867 & 1,061 (18) & -10.82 \\
\midrule
\multicolumn{7}{l}{\textit{Convex multiplier: $g(q_m)=1+q_m^2$}} \\
\multicolumn{7}{l}{\quad Strong-reference mean cost: 1,190} \\
Zero Router & --- & 96.353 & 3.092 & 0.555 & 1,154 (12) & -3.01 \\
$k$NN & With & 95.621 & 3.516 & 0.863 & 1,111 (14) & -6.59 \\
CRATER & With & 95.125 & 4.064 & 0.811 & 1,044 (16) & -12.20 \\
$k$NN & Without & 95.258 & 3.971 & 0.771 & 1,035 (15) & -13.00 \\
CRATER & Without & 95.789 & 3.229 & 0.983 & 1,050 (19) & -11.73 \\
\bottomrule
\end{tabular}
% ===== END tables/routing_price_uniform.tex =====

\end{table}

\FloatBarrier
% ===== END routing_price_sensitivity.tex =====

% ===== END appendix_routing.tex =====

% ===== BEGIN appendix_repeated.tex =====
\section{Repeated-sampling predictions}
\subsection{Training and beta-binomial estimation}
\label{app:repeated-details}

Section~\ref{sec:repeated-generation} uses CRATER with dimension reduction from Section~\ref{sec:results}. Its predictions are held fixed when estimating the beta-binomial parameters.

The beta-binomial parameters are estimated jointly across token budgets 128, 256, 512, 1024, 2048 and 4096, with a separate parameter set for each configuration shared across budgets.

For a success count $d$ out of $R=10$ samples, the beta-binomial mass function is
\[
f_{\mathrm{BB}}(d;\mu,\phi)
=\binom{R}{d}
\frac{\operatorname{Be}\{d+\mu\phi,R-d+(1-\mu)\phi\}}
{\operatorname{Be}\{\mu\phi,(1-\mu)\phi\}}.
\]
Let $\mathcal I_{\mathrm{tr}}$ index the training questions and let $\mathcal B$ contain the six budgets. For each configuration, we maximize
\begin{equation}
\sum_{i\in\mathcal I_{\mathrm{tr}}}\sum_{B\in\mathcal B}
\log f_{\mathrm{BB}}\{D_{im}(B);\mu_{im}(B),\phi_m\}
\label{eq:bb-calibration-objective}
\end{equation}
over $(a_m,b_m,\phi_m)$, with the mean given by~\eqref{eq:pass-mean} and the competing-risks predictions fixed. The same response samples contribute counts at several budgets, so this sum of marginal log likelihoods is a composite likelihood. Evaluation uses $k=1,\ldots,10$, within the available sample count.

For the mixture calculation,
$\E\{(1-P)^k\}=\operatorname{Be}(\mu\phi,(1-\mu)\phi+k)/
\operatorname{Be}(\mu\phi,(1-\mu)\phi)$.
As $\phi\to\infty$, the repeated-success prediction reduces to the plug-in formula evaluated at the calibrated mean $\mu$, which need not equal the original prediction $\widehat p$. The function $1-(1-P)^k$ has second derivative $-k(k-1)(1-P)^{k-2}\leq0$ for $k\geq2$. By Jensen's inequality, substituting the mean success probability can therefore overstate average pass@$k$ under heterogeneous success probabilities. For a fixed positive precision, increasing $\mu$ shifts the beta distribution upward in monotone likelihood-ratio order. The positive calibration slope and common precision therefore preserve the nondecreasing budget order inherited from the primary CIF, as well as nondecreasing pass@$k$ in $k$.

\subsection{Complete repeated-sampling results}
Table~\ref{tab:passk-complete} extends Tables~\ref{tab:limited} and~\ref{tab:pass-time-ablation} to $k=1,\ldots,10$ at each of the six token budgets. For each budget--$k$ combination, predictions and empirical estimates are first averaged over evaluation questions and configurations. The mean absolute differences reported in Section~\ref{sec:repeated-generation} then average the 60 absolute differences between these means, using unrounded values.

% ===== BEGIN tables/passk_complete.tex =====

\begingroup
\small
\setlength{\tabcolsep}{6pt}
\newcommand{\closest}[1]{\textcolor[HTML]{4D8067}{\textbf{#1}}}

\begin{longtable}{rrrrrr}
\caption{Complete mean pass@$k$ results for Tables~\ref{tab:limited}
and~\ref{tab:pass-time-ablation}. Entries average over 1387 evaluation
questions and ten configurations. Empirical estimates use ten response
samples per question--configuration pair. Naive plug-in predictions use
each model's single-sample success probabilities; beta-binomial predictions
use separately estimated calibration layers for the two models. Among
the four predictions in each row, green bold entries indicate the
prediction closest to the empirical estimate in absolute difference,
assessed using unrounded values. Ties are highlighted equally.}
\label{tab:passk-complete}\\

\toprule
 & & \multicolumn{2}{c}{Naive plug-in}
   & \multicolumn{2}{c}{Beta-binomial calibrated} \\
\cmidrule(lr){3-4}\cmidrule(lr){5-6}
$k$ & Empirical
    & Without token effects & CRATER
    & Without token effects & CRATER \\
\midrule
\endfirsthead

\multicolumn{6}{c}{\tablename\ \thetable\ (continued)} \\
\toprule
 & & \multicolumn{2}{c}{Naive plug-in}
   & \multicolumn{2}{c}{Beta-binomial calibrated} \\
\cmidrule(lr){3-4}\cmidrule(lr){5-6}
$k$ & Empirical
    & Without token effects & CRATER
    & Without token effects & CRATER \\
\midrule
\endhead

\midrule
\multicolumn{6}{r}{Continued on next page} \\
\endfoot

\bottomrule
\endlastfoot

\multicolumn{6}{l}{\textit{Budget: 128 tokens}} \\*
1  & 0.0014 & 0.1582 & 0.0113 & 0.0433 & \closest{0.0066} \\*
2  & 0.0020 & 0.2822 & 0.0222 & 0.0548 & \closest{0.0089} \\*
3  & 0.0024 & 0.3816 & 0.0327 & 0.0614 & \closest{0.0103} \\*
4  & 0.0027 & 0.4628 & 0.0429 & 0.0661 & \closest{0.0112} \\*
5  & 0.0029 & 0.5299 & 0.0527 & 0.0696 & \closest{0.0120} \\*
6  & 0.0031 & 0.5860 & 0.0622 & 0.0725 & \closest{0.0126} \\*
7  & 0.0032 & 0.6335 & 0.0714 & 0.0749 & \closest{0.0131} \\*
8  & 0.0033 & 0.6739 & 0.0804 & 0.0769 & \closest{0.0136} \\*
9  & 0.0034 & 0.7085 & 0.0892 & 0.0787 & \closest{0.0140} \\*
10 & 0.0035 & 0.7385 & 0.0977 & 0.0803 & \closest{0.0143} \\

\multicolumn{6}{l}{\textit{Budget: 256 tokens}} \\*
1  & 0.0556 & 0.2803 & 0.0723 & 0.1334 & \closest{0.0530} \\*
2  & 0.0708 & 0.4603 & 0.1311 & 0.1616 & \closest{0.0684} \\*
3  & 0.0797 & 0.5835 & 0.1810 & 0.1771 & \closest{0.0772} \\*
4  & 0.0860 & 0.6714 & 0.2244 & 0.1876 & \closest{0.0834} \\*
5  & 0.0908 & 0.7361 & 0.2628 & 0.1956 & \closest{0.0882} \\*
6  & 0.0947 & 0.7850 & 0.2971 & 0.2020 & \closest{0.0920} \\*
7  & 0.0979 & 0.8227 & 0.3281 & 0.2073 & \closest{0.0952} \\*
8  & 0.1007 & 0.8522 & 0.3563 & 0.2118 & \closest{0.0980} \\*
9  & 0.1031 & 0.8757 & 0.3822 & 0.2157 & \closest{0.1004} \\*
10 & 0.1052 & 0.8946 & 0.4059 & 0.2192 & \closest{0.1026} \\

\multicolumn{6}{l}{\textit{Budget: 512 tokens}} \\*
1  & 0.3853 & 0.4545 & \closest{0.3990} & 0.3310 & 0.3679 \\*
2  & 0.4324 & 0.6655 & 0.5732 & 0.3776 & \closest{0.4221} \\*
3  & 0.4560 & 0.7801 & 0.6745 & 0.4019 & \closest{0.4503} \\*
4  & 0.4713 & 0.8481 & 0.7411 & 0.4180 & \closest{0.4688} \\*
5  & 0.4825 & 0.8913 & 0.7881 & 0.4300 & \closest{0.4824} \\*
6  & 0.4913 & 0.9199 & 0.8229 & 0.4393 & \closest{0.4931} \\*
7  & 0.4985 & 0.9396 & 0.8496 & 0.4470 & \closest{0.5018} \\*
8  & 0.5045 & 0.9535 & 0.8706 & 0.4535 & \closest{0.5092} \\*
9  & 0.5098 & 0.9637 & 0.8874 & 0.4592 & \closest{0.5155} \\*
10 & 0.5144 & 0.9712 & 0.9012 & 0.4641 & \closest{0.5210} \\

\multicolumn{6}{l}{\textit{Budget: 1024 tokens}} \\*
1  & 0.7169 & 0.6499 & 0.7389 & 0.6001 & \closest{0.7229} \\*
2  & 0.7659 & 0.8377 & 0.8788 & 0.6474 & \closest{0.7694} \\*
3  & 0.7882 & 0.9130 & 0.9309 & 0.6707 & \closest{0.7916} \\*
4  & 0.8020 & 0.9489 & 0.9561 & 0.6858 & \closest{0.8056} \\*
5  & 0.8118 & 0.9680 & 0.9702 & 0.6967 & \closest{0.8155} \\*
6  & 0.8192 & 0.9789 & 0.9787 & 0.7052 & \closest{0.8231} \\*
7  & 0.8252 & 0.9856 & 0.9842 & 0.7120 & \closest{0.8292} \\*
8  & 0.8301 & 0.9898 & 0.9880 & 0.7177 & \closest{0.8343} \\*
9  & 0.8344 & 0.9926 & 0.9906 & 0.7226 & \closest{0.8386} \\*
10 & 0.8381 & 0.9945 & 0.9925 & 0.7269 & \closest{0.8423} \\

\multicolumn{6}{l}{\textit{Budget: 2048 tokens}} \\*
1  & 0.8409 & 0.8050 & 0.8463 & 0.8101 & \closest{0.8376} \\*
2  & 0.8848 & 0.9345 & 0.9459 & 0.8421 & \closest{0.8724} \\*
3  & 0.9028 & 0.9717 & 0.9744 & 0.8573 & \closest{0.8883} \\*
4  & 0.9131 & 0.9857 & 0.9859 & 0.8667 & \closest{0.8981} \\*
5  & 0.9200 & 0.9920 & 0.9914 & 0.8735 & \closest{0.9049} \\*
6  & 0.9252 & 0.9952 & 0.9944 & 0.8786 & \closest{0.9101} \\*
7  & 0.9292 & 0.9969 & 0.9961 & 0.8828 & \closest{0.9142} \\*
8  & 0.9325 & 0.9979 & 0.9972 & 0.8862 & \closest{0.9175} \\*
9  & 0.9353 & 0.9986 & 0.9979 & 0.8891 & \closest{0.9204} \\*
10 & 0.9376 & 0.9990 & 0.9984 & 0.8916 & \closest{0.9228} \\

\multicolumn{6}{l}{\textit{Budget: 4096 tokens}} \\*
1  & 0.8733 & 0.8832 & \closest{0.8705} & 0.9022 & 0.8635 \\*
2  & 0.9136 & 0.9688 & 0.9587 & \closest{0.9225} & 0.8950 \\*
3  & 0.9292 & 0.9880 & 0.9817 & \closest{0.9318} & 0.9092 \\*
4  & 0.9379 & 0.9943 & 0.9903 & \closest{0.9374} & 0.9178 \\*
5  & 0.9437 & 0.9969 & 0.9943 & \closest{0.9414} & 0.9238 \\*
6  & 0.9479 & 0.9982 & 0.9964 & \closest{0.9444} & 0.9283 \\*
7  & 0.9512 & 0.9988 & 0.9976 & \closest{0.9468} & 0.9319 \\*
8  & 0.9538 & 0.9992 & 0.9983 & \closest{0.9488} & 0.9348 \\*
9  & 0.9560 & 0.9994 & 0.9988 & \closest{0.9504} & 0.9372 \\*
10 & 0.9578 & 0.9996 & 0.9991 & \closest{0.9519} & 0.9393 \\

\end{longtable}
\endgroup
% ===== END tables/passk_complete.tex =====

\subsection{Sensitivity to calibration parameter sharing}
\label{app:calibration-sensitivity}

The calibration model in~\eqref{eq:pass-mean} uses one set of mean-link parameters $(a_m,b_m)$ and one precision $\phi_m$ for all six budgets within each configuration. This sharing pools information across budgets, but a single calibration curve need not describe the success-count distributions equally well at every budget. At $k=R=10$, the repeated-success prediction is
\[
\widehat\pi_{im}^{(10)}(B)
=1-f_{\mathrm{BB}}\{0;\widehat\mu_{im}(B),\widehat\phi_m\}.
\]
Thus the shortfall at 4096 tokens corresponds to assigning too much probability to zero successes in ten samples. A model can predict the average single-sample success rate accurately and still misestimate the probability that all ten samples fail.

To assess sensitivity to parameter sharing, we keep CRATER's single-sample predictions fixed and compare four calibration specifications. Starting from the shared model, we allow the mean-link parameters to vary by budget, the precision to vary by budget, or both. Let $B_j$ denote the $j$th budget, with $(B_1,\ldots,B_6)=(128,256,512,1024,2048,4096)$. The most flexible specification assigns a separate parameter triple $(a_{mj},b_{mj},\phi_{mj})$ to each configuration--budget pair:
\[
\logit\mu_{im}(B_j)
=a_{mj}+\exp(b_{mj})\logit\widehat p_{im}(B_j),
\qquad \phi_{mj}>0,\quad j=1,\ldots,6.
\]
This gives six parameter triples, or 18 calibration parameters, per configuration. The other specifications share $(a_{mj},b_{mj})$, $\phi_{mj}$, or both across $j$. For a fixed mean, a larger precision implies less residual variation in the latent success probability. In each specification, all calibration parameters are reestimated by the composite likelihood in~\eqref{eq:bb-calibration-objective}, with the appropriate budget indices. The fits use the same 1400 training questions, ten samples per question--configuration pair and six budgets as before. The competing-risks parameters remain fixed, and evaluation outcomes are excluded from estimation.

\begin{table}[!htbp]
\centering
\caption{Sensitivity to sharing calibration parameters across token budgets. ``Shared'' denotes one parameter or parameter pair per configuration across all six budgets; ``By budget'' denotes a separate value or pair at each budget. Parameter counts are per configuration. Mean pass@10 is evaluated at 4096 tokens, where the empirical estimate is 0.9578. All evaluation means average over 1387 questions and ten configurations. The final column reports the average absolute difference (AAD) between predicted and empirical mean pass@$k$ over the 60 combinations of six token budgets and $k=1,\ldots,10$.}
\label{tab:calibration-sharing}
% ===== BEGIN tables/passk_calibration_sensitivity.tex =====
\begin{tabular}{llrrr}
\toprule
Mean link $(a,b)$ & Precision $\phi$ & Parameters & Mean pass@10 & AAD \\
\midrule
Shared    & Shared    &  3 & 0.9393 & 0.00887 \\
By budget & Shared    & 13 & 0.9435 & 0.00772 \\
Shared    & By budget &  8 & 0.9511 & 0.00537 \\
By budget & By budget & 18 & \textcolor[HTML]{4D8067}{\textbf{0.9556}} & 0.00222 \\
\bottomrule
\end{tabular}
% ===== END tables/passk_calibration_sensitivity.tex =====

\end{table}

Table~\ref{tab:calibration-sharing} shows that the maximum-budget discrepancy is sensitive to the calibration specification, particularly the restriction on precision. Because all parameters are reestimated in each fit, the changes cannot be attributed to an isolated change in the numerical value of $\phi$.

The additional flexibility has a cost. The shared specification preserves nondecreasing predictions in budget, as discussed in Appendix~\ref{app:repeated-details}; separate calibration parameters at each budget do not provide this guarantee. Among evaluation comparisons over questions, configurations, adjacent budgets and values of $k$, predicted decreases greater than $10^{-8}$ occur in 8.29\%, 0.15\% and 4.29\% of comparisons for the mean-only, precision-only and fully varying specifications, respectively, compared with none for the shared model. We therefore retain the shared specification in the previous analysis.

\section*{Data Availability Statement} The data, code and notebooks needed to reproduce the results reported in this paper  are available in the repository at \url{https://anonymous.4open.science/r/llm-competing-risks-4463}.
% ===== END appendix_repeated.tex =====

% ===== END appendix.tex =====

\end{document}